\input harvmac.tex
\input epsf


\def\ndt{\noindent}

\def\Ga{{\Gamma}}

\def\K3{{\bf K3}}
\def\journal#1&#2(#3){\unskip, \sl #1\ \bf #2 \rm(19#3) }
\def\andjournal#1&#2(#3){\sl #1~\bf #2 \rm (19#3) }

\def\bar{\overline}
\def\hat{\widehat}

\def\tilde{\widetilde}

\def\frac#1#2{{#1\over#2}}

\def\ket#1{|#1\rangle}

\def\vev#1{\langle#1\rangle}

\def\inbar{\,\vrule height1.5ex width.4pt depth0pt}
\def\IC{\relax\hbox{$\inbar\kern-.3em{\rm C}$}}
\def\IR{\relax{\rm I\kern-.18em R}}
\def\IP{\relax{\rm I\kern-.18em P}}

%
%


%
\catcode`\@=11
\def\slash#1{\mathord{\mathpalette\c@ncel{#1}}}
\overfullrule=0pt

\def\NN{{\cal N}}
\def\OO{{\cal O}}

\def\RR{{\cal R}}

\def\lam{\lambda}

\def\underrel#1\over#2{\mathrel{\mathop{\kern\z@#1}\limits_{#2}}}

\catcode`\@=12


%

\def\ket#1{\left| #1\right\rangle}
\def\vev#1{\left\langle #1 \right\rangle}

\def\tr{{\rm tr}}

\def \sinh{{\rm sinh}}
\def \cosh{{\rm cosh}}


\def \ov {\over}
\def \p {\partial}
\def \ha {{1 \ov 2}}
\def \al {\alpha}
\def \lam {\lambda}
\def \Lam {\Lambda}
\def \sig {\sigma}
\def \Sig {\Sigma}
\def \om {\omega}
\def \Om {\Omega}
\def \ep {\epsilon}

\def \ga {\gamma}
\def \apr {\alpha'}

\def\le{\left}
\def\ri{\right}

\def\th{\theta}

\def\IL{\relax{\rm I\kern-.18em L}}
\def\IH{\relax{\rm I\kern-.18em H}}
\def\IR{\relax{\rm I\kern-.18em R}}
\def\IC{\relax\hbox{$\inbar\kern-.3em{\rm C}$}}
\def\IZ{{\bf Z}}





\def\makeblankbox#1#2{\hbox{\lower\dp0\vbox{\hidehrule{#1}{#2}%
   \kern -#1
   \hbox to \wd0{\hidevrule{#1}{#2}%
      \raise\ht0\vbox to #1{}
      \lower\dp0\vtop to #1{}
      \hfil\hidevrule{#2}{#1}}%
   \kern-#1\hidehrule{#2}{#1}}}%
}%
\def\hidehrule#1#2{\kern-#1\hrule height#1 depth#2 \kern-#2}%
\def\hidevrule#1#2{\kern-#1{\dimen0=#1\advance\dimen0 by #2\vrule
    width\dimen0}\kern-#2}%
\def\openbox{\ht0=1.2mm \dp0=1.2mm \wd0=2.4mm  \raise 2.75pt
\makeblankbox {.25pt} {.25pt}  }

\def\bun#1/#2{\leavevmode
   \kern.1em \raise .5ex \hbox{\the\scriptfont0 #1}%
   \kern-.1em $/$%
   \kern-.15em \lower .25ex \hbox{\the\scriptfont0 #2}%
}

\def\opensquare{\ht0=3.4mm \dp0=3.4mm \wd0=6.8mm  \raise 2.7pt
\makeblankbox {.25pt} {.25pt}  }


\def\sector#1#2{\ {\scriptstyle #1}\hskip 1mm
\mathop{\opensquare}\limits_{\lower
1mm\hbox{$\scriptstyle#2$}}\hskip 1mm}

\def\tsector#1#2{\ {\scriptstyle #1}\hskip 1mm
\mathop{\opensquare}\limits_{\lower
1mm\hbox{$\scriptstyle#2$}}^\sim\hskip 1mm}

\def\IZ{{\bf Z}}

\def\De{\Delta}
\def\bep{\bar \ep}


\lref\LV{Landsman and Van Weert , Phys.\ Rept.\  }

\lref\sred{M.~Srednicki, Phys.\ Rev.\ E {\bf 50} 888 (1994).}

\lref\ccgi{G.~Casati1, B.~V.~Chirikov1,2, I.~Guarneri1 and
F.~M.~Izrailev, cond-mat/9607081.}

\lref\fcic{Y.~V.~Fyodorov, O.~A.~Chubykalo, F~.M~Izrailev and
G.~Casati, Phys. \ Rev. \ Lett. \ {\bf 76} 1603 (1996).}

\lref\HorowitzPQ{
  G.~T.~Horowitz and S.~F.~Ross,
  ``Possible resolution of black hole singularities from large N gauge
  theory,''
  JHEP {\bf 9804}, 015 (1998)
  [arXiv:hep-th/9803085].
}

\lref\gkp{ S.~S.~Gubser, I.~R.~Klebanov and A.~M.~Polyakov,
``Gauge theory correlators from non-critical string theory,''
Phys.\ Lett.\ B {\bf 428}, 105 (1998) [arXiv:hep-th/9802109].
}

\lref\MaldacenaRE{ J.~M.~Maldacena, ``The large N limit of
superconformal field theories and supergravity,'' Adv.\ Theor.\
Math.\ Phys.\  {\bf 2}, 231 (1998) [Int.\ J.\ Theor.\ Phys.\  {\bf
38}, 1113 (1999)] [arXiv:hep-th/9711200].
}

\lref\witten{ E.~Witten, ``Anti-de Sitter space and holography,''
Adv.\ Theor.\ Math.\ Phys.\  {\bf 2}, 253 (1998)
[arXiv:hep-th/9802150].
}

\lref\maldat{ J.~M.~Maldacena, ``Eternal black holes in
Anti-de-Sitter,'' arXiv:hep-th/0106112.
}

\lref\witt{ E.~Witten, ``Anti-de Sitter space, thermal phase
transition, and confinement in gauge theories,'' Adv.\ Theor.\
Math.\ Phys.\  {\bf 2}, 505 (1998) [arXiv:hep-th/9803131].
}

\lref\sussWi{
  L.~Susskind and E.~Witten,
  ``The holographic bound in anti-de Sitter space,''
  arXiv:hep-th/9805114.
}

\lref\shenker{
  L.~Fidkowski, V.~Hubeny, M.~Kleban and S.~Shenker,
  ``The black hole singularity in AdS/CFT,''
  JHEP {\bf 0402}, 014 (2004)
  [arXiv:hep-th/0306170].
}

\lref\mot{
  L.~Motl and A.~Neitzke,
  ``Asymptotic black hole quasinormal frequencies,''
  Adv.\ Theor.\ Math.\ Phys.\  {\bf 7}, 307 (2003)
  [arXiv:hep-th/0301173].
}

\lref\ricar{
  V.~Cardoso, J.~Natario and R.~Schiappa,
  ``Asymptotic quasinormal frequencies for black holes in non-asymptotically
  flat spacetimes,''
  J.\ Math.\ Phys.\  {\bf 45}, 4698 (2004)
  [arXiv:hep-th/0403132].
}

\lref\FLlong{G.~Festuccia and H.~Liu, to appear.}

\lref\horomal{
  G.~T.~Horowitz and J.~Maldacena,
  ``The black hole final state,''
  JHEP {\bf 0402}, 008 (2004)
  [arXiv:hep-th/0310281].
}

\lref\HertogRZ{
  T.~Hertog and G.~T.~Horowitz,
  ``Towards a big crunch dual,''
  JHEP {\bf 0407}, 073 (2004)
  [arXiv:hep-th/0406134].
}

\lref\HertogHU{
  T.~Hertog and G.~T.~Horowitz,
  ``Holographic description of AdS cosmologies,''
  arXiv:hep-th/0503071.
}

\lref\Fyodorov{
  A.D. Mirlin, Yan Fyodorov,
  ``Universality of level correlation functions of sparse random matrices''
  J.Phys. A: Math. Gen. {\bf 24} (1991) 2273-2286}

\lref\Rodgers{
  G.J. Rodgers, A.J. Bray,
  ``Density of states of a sparse random matrices''
  PRB {\bf 37, 7} (1988) 3557}

\lref\Semerjian{
  G. Semerjian et al,
  ``Sparse random matrices: the eigenvalue spectrum revisited''
  J. Phys. A: Math. Gen. {\bf 35} (2002) ( 4837-4851}

\lref\barbon{
  J.~L.~F.~Barbon and E.~Rabinovici,
  ``Topology change and unitarity in quantum black hole dynamics,'' hep-th/0503144;
  J.~L.~F.~Barbon and E.~Rabinovici,
  ``Long time scales and eternal black holes,''
  Fortsch.\ Phys.\  {\bf 52}, 642 (2004)
  hep-th/0403268; J.~L.~F.~Barbon and E.~Rabinovici,
  ``Very long time scales and black hole thermal equilibrium,''
  JHEP {\bf 0311}, 047 (2003)
 hep-th/0308063.
}

\lref\banks{
  T.~Banks and W.~Fischler,
  ``Space-like singularities and thermalization,''
  arXiv:hep-th/0606260.
}

\lref\carter{
  J.~M.~Bardeen, B.~Carter and S.~W.~Hawking,
  ``The Four laws of black hole mechanics,''
  Commun.\ Math.\ Phys.\  {\bf 31}, 161 (1973).
}

\lref\hawking{
  S.~W.~Hawking,
  ``Particle Creation By Black Holes,''
  Commun.\ Math.\ Phys.\  {\bf 43}, 199 (1975)
  [Erratum-ibid.\  {\bf 46}, 206 (1976)].
}

\lref\bekenstein{
  J.~D.~Bekenstein,
  ``Black holes and entropy,''
  Phys.\ Rev.\ D {\bf 7}, 2333 (1973);
   J.~D.~Bekenstein,
  ``Generalized second law of thermodynamics in black hole physics,''
  Phys.\ Rev.\ D {\bf 9}, 3292 (1974).
}

\lref\Minwa{
  O.~Aharony, J.~Marsano, S.~Minwalla, K.~Papadodimas and M.~Van Raamsdonk,
   ``A first order deconfinement transition in large N Yang-Mills theory on  a
  small $S^3$,''
  Phys.\ Rev.\ D {\bf 71}, 125018 (2005)
  [arXiv:hep-th/0502149].
}

\lref\minW{
  O.~Aharony, J.~Marsano, S.~Minwalla, K.~Papadodimas and M.~Van Raamsdonk,
   ``The Hagedorn / deconfinement phase transition in weakly coupled large N
  gauge theories,''
  Adv.\ Theor.\ Math.\ Phys.\  {\bf 8}, 603 (2004)
  [arXiv:hep-th/0310285].
}

\lref\klose{
  N.~w.~Kim, T.~Klose and J.~Plefka,
  ``Plane-wave matrix theory from N = 4 super Yang-Mills on R x S**3,''
  Nucl.\ Phys.\ B {\bf 671}, 359 (2003)
  [arXiv:hep-th/0306054].
}

\lref\HorowitzHE{
  G.~T.~Horowitz and J.~M.~Maldacena,
  JHEP {\bf 0402}, 008 (2004)
  [arXiv:hep-th/0310281].
}

\lref\LoweXM{
  D.~A.~Lowe and L.~Thorlacius,
  Phys.\ Rev.\ D {\bf 73}, 104027 (2006)
  [arXiv:hep-th/0601059].
}

\lref\vijayg{
  V.~Balasubramanian, D.~Marolf and M.~Rozali,
  arXiv:hep-th/0604045.
}

\lref\hawkingpage{
  S.~W.~Hawking and D.~N.~Page,
  Commun.\ Math.\ Phys.\  {\bf 87}, 577 (1983).
}

\lref\yaffe{
  L.~G.~Yaffe,
  Rev.\ Mod.\ Phys.\  {\bf 54}, 407 (1982).
}

\lref\guidoliu{
  G.~Festuccia and H.~Liu,
   ``Excursions beyond the horizon: Black hole singularities in Yang-Mills
  JHEP {\bf 0604}, 044 (2006)
  [arXiv:hep-th/0506202].
}

\lref\brigante{
  M.~Brigante, G.~Festuccia and H.~Liu,
   ``Inheritance principle and non-renormalization theorems at finite
  Phys.\ Lett.\ B {\bf 638}, 538 (2006)
  [arXiv:hep-th/0509117].
}

\lref\semenoff{
  A.~J.~Niemi and G.~W.~Semenoff,
  Annals Phys.\  {\bf 152}, 105 (1984).
}

\lref\thorn{
  C.~B.~Thorn,
   ``Infinite N(C) QCD At Finite Temperature: Is There An Ultimate
  Temperature?,''
  Phys.\ Lett.\ B {\bf 99}, 458 (1981).
}

\lref\penrose{
  R.~Penrose,
  ``Singularities And Time Asymmetry,'' in General Relativity:
  An Einstein Centenary Survey, edited by S. W. Hawking and W. Israel
  (Cambridge Univ. Press, Cambridge, England, 1979).
}

\lref\arnold{
  V.I.~Arnold, A.~Avez,
  ``Ergodic problems of classical mechanics,''
Redwood City, Calif. : Addison-Wesley, the Advanced Book Program,
c1989.}

\lref\bellac{
  M.~Le Bellac, ``Thermal field theory,''  Cambridge, New York:
Cambridge University Press, 1996.
 }

\lref\wigner{
  E.~Wigner,
``Characteristics Vectors of Bordered Matrices with Infinite
Dimensions,''
 Ann.Math. {\bf 62} 548 (1955), {\bf 65} 203 (1957),
}

\lref\casati{ Fyodorov, Y. V. and Chubykalo, O A. and Izrailev, F.
M and Casati, G.,
  ``Wigner Random Banded Matrices with Sparse Structure: Local Spectral Density of States,''
  Phys. Rev. Lett. {\bf 76}, 1603 (1996)
}

\lref\maldsuss{
  J.~M.~Maldacena and L.~Susskind,
  Nucl.\ Phys.\ B {\bf 475}, 679 (1996)
  [arXiv:hep-th/9604042].
}

\lref\vjay{
  V.~Balasubramanian, P.~Kraus and M.~Shigemori,
   ``Massless black holes and black rings as effective geometries of the D1-D5
  system,''
  Class.\ Quant.\ Grav.\  {\bf 22}, 4803 (2005)
  [arXiv:hep-th/0508110].
}

\lref\thooft{
  G.~'t Hooft,
  Nucl.\ Phys.\ B {\bf 538}, 389 (1999)
  [arXiv:hep-th/9808113].
}

\lref\gross{
  D.~J.~Gross and A.~Neveu,
  Phys.\ Rev.\ D {\bf 10}, 3235 (1974).
}

\lref\lautrup{
  B.~Lautrup,
  Phys.\ Lett.\ B {\bf 69}, 109 (1977).
}

\lref\tHooftRE{G.~'t Hooft,  in The Whys of subnuclear physics,
Proc. Int. School, Erice, Italy, 1977, ed. A. Zichichi (Plenum,
New York, 1978.}

\lref\beneke{
  M.~Beneke,
  Phys.\ Rept.\  {\bf 317}, 1 (1999)
  [arXiv:hep-ph/9807443].
}

\lref\GrossBR{
  D.~J.~Gross, R.~D.~Pisarski and L.~G.~Yaffe,
  Rev.\ Mod.\ Phys.\  {\bf 53}, 43 (1981).
}

\lref\Solone{
  D.~Birmingham, I.~Sachs and S.~N.~Solodukhin,
  ``Relaxation in conformal field theory, Hawking-Page transition, and
  quasinormal/normal modes,''
  Phys.\ Rev.\  D {\bf 67}, 104026 (2003)
  [arXiv:hep-th/0212308].
  S.~N.~Solodukhin,
  ``Can black hole relax unitarily?,''
  arXiv:hep-th/0406130.
  S.~N.~Solodukhin,
  ``Restoring unitarity in BTZ black hole,''
  Phys.\ Rev.\  D {\bf 71}, 064006 (2005)
  [arXiv:hep-th/0501053].
}

\lref\miaoli{
  M.~Li,
   ``Evidence for large N phase transition in N = 4 super Yang-Mills theory  at
  finite temperature,''
  JHEP {\bf 9903}, 004 (1999)
  [arXiv:hep-th/9807196];
  Y.~h.~Gao and M.~Li,
   ``Large N strong/weak coupling phase transition and the correspondence
  principle,''
  Nucl.\ Phys.\ B {\bf 551}, 229 (1999)
  [arXiv:hep-th/9810053].
}

\lref\kumar{
  S.~A.~Hartnoll and S.~Prem Kumar,
  JHEP {\bf 0512}, 036 (2005)
  [arXiv:hep-th/0508092].
}

\lref\hubeny{
  V.~E.~Hubeny, H.~Liu and M.~Rangamani,
  arXiv:hep-th/0610041.
}

\lref\peres{
  A.~Peres,
  Phys. \ Rev. \ A {\bf 30} (1984) 504.
}

\lref\srednicki{
  M.~Sredicki,
  J. \ Phys. \ A {\bf 32} (1999) 1163; cond-mat/9809360.
}

\lref\bkl{
  V.~A.~Belinskii, I.~M.~Khalatnikov, and E.~M.~Lifshitz,
  Adv. \ Phys. \
  {\bf 19}, 525 (1970).
}

\lref\MathurZP{
  S.~D.~Mathur,
  Fortsch.\ Phys.\  {\bf 53}, 793 (2005)
  [arXiv:hep-th/0502050].
}

\lref\KlebanRX{
  M.~Kleban, M.~Porrati and R.~Rabadan,
  JHEP {\bf 0410}, 030 (2004)
  [arXiv:hep-th/0407192].
}

\lref\furuuch{
  K.~Furuuchi,
  Phys.\ Rev.\ D {\bf 72}, 066009 (2005)
  [arXiv:hep-th/0505148].
}

\lref\sundborg{
  B.~Sundborg,
  Nucl.\ Phys.\ B {\bf 573}, 349 (2000)
  [arXiv:hep-th/9908001].
}

\lref\adam{
  A.~Donovan, G.~Festuccia and H.~Liu, in progress.
}

\lref\minahan{
 See e.g.  J.~A.~Minahan and K.~Zarembo,
  JHEP {\bf 0303}, 013 (2003)
  [arXiv:hep-th/0212208];
  N.~Beisert, C.~Kristjansen and M.~Staudacher,
  Nucl.\ Phys.\ B {\bf 664}, 131 (2003)
  [arXiv:hep-th/0303060].
  I.~Bena, J.~Polchinski and R.~Roiban,
  Phys.\ Rev.\ D {\bf 69}, 046002 (2004)
  [arXiv:hep-th/0305116].
  V.~A.~Kazakov, A.~Marshakov, J.~A.~Minahan and K.~Zarembo,
  JHEP {\bf 0405}, 024 (2004)
  [arXiv:hep-th/0402207].
  M.~Staudacher,
  JHEP {\bf 0505}, 054 (2005)
  [arXiv:hep-th/0412188].
  N.~Beisert and M.~Staudacher,
  Nucl.\ Phys.\ B {\bf 727}, 1 (2005)
  [arXiv:hep-th/0504190].
}

\lref\KalyanaRamaZJ{
  S.~Kalyana Rama and B.~Sathiapalan,
  Mod.\ Phys.\ Lett.\ A {\bf 14}, 2635 (1999)
  [arXiv:hep-th/9905219].
}

\lref\BiroSH{
  T.~S.~Biro, C.~Gong and B.~Muller,
  Phys.\ Rev.\ D {\bf 52}, 1260 (1995)
  [arXiv:hep-ph/9409392];
  B.~Muller and A.~Trayanov,
  Phys.\ Rev.\ Lett.\  {\bf 68}, 3387 (1992).
 T.~S.~Biro, M.~Feurstein and H.~Markum,
  Heavy Ion Phys.\  {\bf 7}, 235 (1998)
  [arXiv:hep-lat/9711002].
}

\lref\SrednickiFE{
  M.~Srednicki,
  arXiv:hep-th/0207090.
}




\Title{\vbox{\baselineskip12pt \hbox{hep-th/0611098}
\hbox{MIT-CTP-3783}
}}%
 {\vbox{\centerline{The arrow of time, black holes,
 }
 \smallskip
 \medskip
 \centerline{and quantum mixing of large $N$ Yang-Mills theories}
 } }

\smallskip
\centerline{Guido Festuccia and Hong Liu }
\medskip

\centerline{\it  Center for Theoretical Physics} \centerline{\it
Massachusetts Institute of Technology} \centerline{\it Cambridge,
Massachusetts, 02139}

\smallskip

\smallskip

\smallskip

\vglue .3cm

\bigskip
\noindent

Quantum gravity in an AdS spacetime is
described by an $SU(N)$ Yang-Mills
theory on a sphere, a bounded many-body
system. We argue that in the high
temperature phase the theory is
intrinsically non-perturbative in the
large $N$ limit. At any nonzero value
of the 't Hooft coupling $\lam$, an
exponentially large (in $N^2$) number
of free theory states of wide energy
range (or order $N$) mix under the
interaction. As a result the planar
perturbation theory breaks down.
 We argue that an arrow of time emerges
 and the dual string configuration should be interpreted as a
stringy black hole.


 \Date{Oct. 27, 2006}


\bigskip



\newsec{Introduction}

While the equations of general relativity are time symmetric
themselves, one often finds solutions with an intrinsic arrow of
time, due to the presence of spacelike singularities. Familiar
examples include FRW cosmologies and the formation of a black hole
in a gravitational collapse. In the case of a gravitational
collapse to form a black hole, the direction of time appears to be
thermodynamic, since a black hole behaves like a thermodynamical
system~\refs{\carter, \bekenstein,\hawking}. It has also long been
speculated that the thermodynamic arrow of time observed in nature
may be related to the big bang singularity~\refs{\penrose}.

In an anti-de Sitter (AdS) spacetime, a microscopic understanding
of the emergence of thermodynamic behavior in a gravitational
collapse can be achieved using  the AdS/CFT
correspondence~\refs{\MaldacenaRE,\witten,\gkp}, which states that
quantum gravity in an asymptotic $AdS_5 \times S^5$ spacetime is
described by an $\NN=4$ $SU(N)$ super-Yang-Mills (SYM) theory on
an $S^3$.

The classical gravity limit of the AdS string theory corresponds
to the large $N$ and large 't Hooft coupling limit of the
Yang-Mills theory. A matter distribution of classical mass $M$ in
AdS,
 can be identified\foot{See Appendix A for a brief review of
 parameter translations in AdS/CFT.} with an excited state of energy $E = \mu N^2$
 in the SYM theory with $\mu$ a constant independent of
 $N$. The gravitational collapse of the  matter distribution can be
 identified with the thermalization of the
 corresponding state in SYM theory, with the resulting black
 hole\foot{Assume $M$ is sufficiently big that a big black
holes in AdS is formed, which also implies that $\mu$ should be
sufficiently big.} identified with thermal
equilibrium~\refs{\witt,\maldat}. In this context, it is natural
to suspect that the appearance of a spacelike singularity at the
end point of a collapse should be related to certain aspect of
thermalization in the SYM theory\foot{Some interesting ideas
regarding spacelike singularities and thermalization have also
been discussed recently in~\refs{\banks}.}

A crucial element in the above description is the large $N$ limit.
$\NN=4$ SYM theory on $S^3$ is a closed, bounded quantum
mechanical system with a discrete energy spectrum. At any finite
$N$, no matter how large, such a theory is quasi-periodic
(i.e.~has recurrences), time reversible, and never really
thermalizes.  However, to match the picture of a gravitational
collapse in classical gravity, an arrow of time should emerge in
the large $N$ limit for the SYM theory in a generic state of
energy $E = \mu N^2$ with a sufficiently large $\mu$. This
consistency requirement immediately raises several questions:

\item{1.} What is the underlying physical mechanism for the
emergence of an arrow of time in Yang-Mills theory?

\item{2.} Is large 't Hooft coupling needed?

\item{3.} Suppose an arrow of time also emerges at small 't Hooft
coupling, what would be the bulk string theory interpretation of
the SYM theory in such an excited state? A stringy black hole?
Does such a stringy black hole have a singularity?

\item{4.} Is there a large $N$ phase transition as one decreases
the 't Hooft coupling from infinity to zero?

\ndt It would be very desirable to have a clear physical
understanding of the above questions, which could shed light on
how spacelike singularities appear in the classical limit of a
quantum gravity and thus lead to an understanding of their
resolution in a quantum theory.

The emergence of an arrow of time is also closely related to the
information loss paradox\foot{To our knowledge this connection was
first pointed out in the context of AdS/CFT in~\refs{\maldat}.}.
At finite $N$, the theory is unitary and there is no information
loss. But in the large $N$ limit, an arrow of time emerges and the
information is lost, since one cannot recover the initial state
from the final thermal equilibrium. Thus the information loss in a
gravitational collapse is clearly a consequence of the classical
approximation (large $N$ limit), but not a property of the full
quantum theory. While AdS/CFT in principle resolves the
information loss paradox, it remains a puzzle whether one can
recover the lost information using a semi-classical
reasoning\foot{See
e.g.~\refs{\maldat,\barbon,\KlebanRX,\HorowitzHE,\MathurZP,\LoweXM,\vijayg,\Solone}
for recent discussions.}. From this perspective it would also be
valuable to understand the various questions listed in the last
paragraph.

 The purpose of the paper is to suggest a simple mechanism for the
emergence of an arrow of time in the large $N$ limit and to
initiate a statistical approach to understanding the quantum
dynamics of a YM theory in highly excited states. In particular,
we argue that the perturbative planar expansions break down for
real-time correlation functions and that there is a large $N$
``phase transition'' at zero 't~Hooft
coupling\foot{See~\refs{\miaoli,\kumar} for some earlier
discussion of a possible large $N$ phase transition in $\lam$.}.
We also argue that time irreversibility occurs for any nonzero
value of the 't Hooft coupling.

The plan of the paper is as follows. In section 2 we introduce the
subject of our study: a family of matrix quantum mechanical
systems including $\NN=4$ SYM on $S^3$. We highlight some relevant
features of the energy spectrum of these theories. Motivated by
the classical mixing properties, we introduce observables which
could signal time irreversibility. The simplest of them are
real-time correlation functions at finite temperature, which
describe non-equilibrium linear responses of the systems. The rest
of the paper is devoted to studying these observables, first in
perturbation theory, and then using a non-perturbative statistical
method. In sec 3 we compute real-time correlation functions in
perturbation theory. We find that at any finite order in
perturbation theory, the arrow of time does not emerge. In sec 4
we argue that the planar perturbative expansion has a zero radius
of convergence and cannot be used to understand the long time
behavior of the system. In section 5 we give a simple physical
explanation for the breakdown of perturbation theory. We argue
that for any nonzero 't Hooft coupling, an exponentially large (in
$N^2$) number of free theory states of wide energy range (or order
$N$) mix under the interaction. As a consequence small $\lam$ and
long time limits do not commute at infinite $N$. In section 6 we
develop a statistical approach to studying the dynamics of the
theories in highly excited states, which indicates that time
irreversibility occur for any nonzero 't Hooft coupling $\lam$. We
conclude in section 7 with a discussion of implications of our
results.

\newsec{Prelude: theories and observables of interest}

In this section we introduce the
systems and observables we want to
study.

\subsec{Matrix mechanical systems}

We consider generic matrix quantum mechanical systems of the form
 \eqn\onedAc{
 S =  N \tr \int\!dt  \;
\biggl[  \sum_\al
 \le(\frac{1}{2} (D_t M_\al)^2
 - \ha \om_{\al}^2 M_\al^2 \ri)
    \, \biggr]  - \int d t \, V (M_\al; \lam)
 }
which satisfy the following
requirements:

\item{1.} $M_\al$ are $N \times N$ matrices and $D_t M_\al = \p_t
- i [A, M_\al]$ is a covariant derivative. One can also include
fermionic matrices, but they will not play an important role in
this paper and for simplicity of notations we suppress them.

\item{2.} The frequencies $\om_\al$ in \onedAc\ are nonzero for
any $\al$, i.e. the theory has a mass gap and a unique vacuum.

\item{3.} The number of matrices is greater than one and can be
infinite. When there is an infinite number of matrices, we require the
theory to be obtainable from a renormalizable field theory on a compact space.

\item{4.} $V (M_\al;\lam)$ can be written as a sum of {\it
single-trace} operators and is controlled by a coupling constant
$\lam$, which remains fixed in the large $N$ limit.

\medskip

$\NN=4$ SYM on $S^3$ is an example of
such systems with an infinite number of
matrices (including fermions) when the
Yang-Mills and matter fields are
expanded in terms of spherical
harmonics on $S^3$ (see
e.g.~\refs{\klose,\Minwa}). In this
case, $\om_\al$ are integer or
half-integer multiples of a fundamental
frequency $\om_0 = 1/R$ with $R$ the
radius of the $S^3$.  The number of
modes with frequencies $\om_\al = {k
\ov R}$ increases with $k$ as a power.
$V(M_\al;\lam)$ can be schematically
written as\foot{The precise form of the
interactions depends on the choice of
gauge. It is convenient to choose
Coulomb gauge $\nabla \cdot \vec A =0$,
in which the longitudinal component of
the gauge field is set to zero. In this
gauge, $M_{\al}$ include also
non-propagating modes coming from
harmonic modes of ghosts and the zero
component of the gauge field.}
 \eqn\higT{
 V =  N \le(\sqrt{\lam}  V_3 (M_\al) + \lam
 V_{4} (M_\al) \ri)
 }
 where $V_{3}$ and $V_{4}$ contain infinite sums of  single-trace operators
 which are cubic and quartic in $M_\al$ and $\p_t M_\al$. $\lam = g_{YM}^2 N$ is the 't Hooft
 coupling.

In this paper we work in the large $N$ limit throughout. Our
discussion will only depend on the large $N$ scaling of various
physical quantities and not on the specific
 structure of the theories in \onedAc\ like the precise field contents
 and exact forms of interactions. For purpose of illustration,
 we will often use as a specific example the following simple system
  \eqn\GonedAc{\eqalign{
 S & =  {N \ov 2} \tr \int\!d t  \;
  \le[  (D_t M_1)^2 +  (D_t M_2)^2
 - \om_0^2 ( M_1^2 + M_2^2)  -   \lam  M_1 M_2 M_1 M_2  \ri] \ .
 }}

\subsec{Energy spectrum}

\onedAc\ has a $U(N)$ gauge symmetry and physical states are
singlets of $U(N)$. One can classify energy eigenstates of a
theory by how their energies scale with $N$ in the large $N$
limit. We will call the sector of states whose energies (as
measured from the vacuum) are of order $O(1)$ the low energy
sector. As motivated in the introduction, we are mainly interested
in the sector of states whose energies are of order $\mu N^2$ with
$\mu$ independent of $N$, which will be called the high energy
sector. The density of states in the low energy sector is of order
$O(1)$, i.e. independent of $N$, while that of the high energy
sector can be written in a form
 \eqn\djd{
 \Om (E) \sim e^{s (\mu) N^2}, \qquad  E = \mu N^2
 }
with $s(\mu)$ some function independent
of $N$. \djd\ follows from the fact
that the number of ways to construct a
state of energy of order $O(N^2)$ from
$O(N^2)$ oscillators of frequency of
$O(1)$ is an exponential in $N^2$. The
presence of interaction should not
change this behavior at least for $\mu$
sufficiently large. \djd\ is the reason
why we restrict to more than one matrix
in \onedAc. For a gauged matrix quantum
mechanics with a single matrix one can
reduce the matrix to its eigenvalues
and \djd\ does not apply.
 When $\mu$ is sufficiently
large, $s(\mu)$ should be a monotonically increasing
function\foot{That is, the theory should have a positive specific
heat for $\mu$ sufficiently large.} of $\mu$ and we will restrict
our definition of high energy sector to such energies.

For $\NN=4$ SYM, states in the low energy sector correspond to
fundamental string states in the AdS spacetime, while the states
in the high energy sectors may be considered as black hole
microstates\foot{Note that at a sufficiently high energy, the most
entropic object in AdS is a big black hole.}.

A convenient way to study a system in excited states is to put it
in a canonical ensemble with a temperature $T= {1 \ov \beta}$. The
partition function and free energy are defined by ($\Tr$ denotes
sum over all physical states and $H$ is the Hamiltonian)
 \eqn\jdis{
 Z= \Tr e^{-\beta H} = e^{-\beta F} \ .
 }
We will always keep $T$ fixed in the
large $N$ limit. Below low and high
temperature refers to how the
temperature is compared with the mass
gap of a theory\foot{For example for
$\NN=4$ SYM on $S^3$, low (high)
temperature means $T \ll {1 \ov R}$ ($T
\gg {1 \ov R}$)}. As one varies $T$,
different parts of the energy spectrum
are probed. For the family of matrix
quantum mechanical systems \onedAc,
there are two distinct temperature
regimes. At low temperature, one probes
the low energy sector and the free
energy $F$ is of order $O(1)$. At high
temperature $F$ is of order $O(N^2)$
and
 the high energy sector is probed. It may seem surprising at
 first sight that one can probe the sector of energies of $O(N^2)$
 using a temperature of $O(1)$. This is due to the
 large entropy factor \djd\ which compensates the Boltzmann
 suppression. For $\NN=4$ SYM theory
at strong coupling, there is a first order phase transition
separating the two
 regimes at a temperature of order
 $1/R$,
 where $R$ is the AdS radius~\refs{\hawkingpage,\witten,\witt}.
A first order phase transition has also been found for various
theories in the family of \onedAc\ at weak
coupling~\refs{\sundborg,\minW,\Minwa}.

An important feature of the high energy sector is that the large
$N$ limit is like a thermodynamic limit with $N^2$ playing the
analogous role of the volume factor. In this limit the number of
degrees of freedom goes to infinity while the average excitations
per degree of freedom remain finite. The thermal partition
function
 \eqn\Diep{
 Z (\beta) = \Tr e^{-\beta H} = \int d E \, \Om (E) e^{-\beta E}
 }
is sharply peaked at an energy $E_\beta \sim O(N^2)$ (with a width
of order $O(N)$) determined by
 \eqn\Diru{
{\p S(E) \ov \p E} \biggr|_{E_\beta} = \beta, \qquad S (E) = \log
\Om (E)
 }
Note that the leading $N$ dependence of $S(E)$ has the form $S (E)
= N^2 s (\mu)$ ~(see \djd) with $\mu = E/N^2$ characterizing the
average excitations per oscillator degree of freedom. Equation
\Diru\ can also be interpreted as the equivalence between
canonical and microcanonical ensemble\foot{In contrast in the low
energy sector, since both the free energy and the density of
states are independent of $N$, in generic models there is no large
parameter that one can use to perform the saddle point
approximation to equate two ensembles.}. Note that since $F \sim
O(N^2)$, the high temperature phase can be considered a
``deconfined'' phase~\refs{\thorn,\witt}.

\subsec{Observables}

In a classical Hamiltonian system,  time irreversibility is
closely related with the mixing property of the system, which can
be stated as follows. Consider time correlation functions
 \eqn\fjsa{
C_{AB} (t) = \vev{A(\Phi^t X) B (X)} - \vev{A}\vev{B}
 }
where $A,B$ are functions on the classical phase space
parameterized by $X$. $\Phi^t X$ describes the Hamiltonian flow,
where $\Phi^t$ is a one-parameter group of volume-preserving
transformations of the phase space onto itself. $\vev{...}$ in
\fjsa\ denotes phase space average over a constant energy surface.
The system is mixing\foot{Note that mixing is a stronger property
than ergodic which involves long time average. The ergodic and
mixing properties can also be characterized in terms of the
spectrum of the Koopman operator. For example,
 a system is mixing iff the eigenvalue $1$ is
simply degenerate and is the only proper eigenvalue of the Koopman
operator~\refs{\arnold}.} iff~\refs{\arnold}
 \eqn\fjjs{
 C_{AB} (t) \to 0, \qquad t \to \infty
 }
for any {\it smooth} $L^2$ functions $A$ and $B$.

 The closest analogue
of \fjsa\ for the matrix quantum mechanical systems we are
considering would be
 \eqn\ffj{
 G_i (t) = \vev{i|\OO(t) \OO(0)|i} - \vev{i|\OO (0)|i}^2
 }
where $\ket{i}$ is a generic energy
eigenstate in the high energy sector,
and $\OO$ is an arbitrary gauge
invariant operator which when acting on
the vacuum creates a state of finite
energy of order $O(1)$. More
explicitly, denoting $\ket{\psi_\OO} =
\OO (0) \ket{\Om}$ with $\ket{\Om}$ the
vacuum, we require $\vev{\psi_\OO|H
|\psi_\OO} \sim O(1)$. Note that for
$\NN=4$ SYM on $S^3$, a {\it local}
operator $O(t,\vec x)$ of dimension
$O(1)$ on $S^3$ is not allowed by this
criterion since $O(t,\vec x)$ creates a
state of infinite energy. To construct
a state of finite energy one can smear
the local operator over a spatial
volume, e.g. by considering operators
with definite angular momentum on
$S^3$. Without loss of generality, we
can take $\OO$ to be of the form
 \eqn\fjd{
 \OO = \tr (M_{\al_1} \cdots M_{\al_{n_1}}) \tr (M_{\beta_{1}}
 \cdots M_{\beta_{n_2}}) \cdots \tr(M_{\ga_{1}}
 \cdots M_{\ga_{n_k}})
 }
with the total number of matrices $K = \sum_{i=1}^k n_k$
independent of $N$. We will call such operators small operators.
The reason for restricting to small operators is that they have a
well defined large $N$ limit in the sense defined
in~\refs{\yaffe}. More explicitly, if one treats the large $N$
limit of a matrix quantum mechanics as a classical system, then
\fjd\ with $K \sim O(1)$ are {\it smooth} functions on the
corresponding classical pase space. From AdS point of view, such
operators correspond to fundamental string probes which do not
deform the background geometry. If for all small operators $\OO$
and generic states $\ket{i}$ in the high energy sector
 \eqn\rhjs{
 G_i (t) \to 0, \qquad t \to \infty
 }
one can say the system develops an
arrow of time. In particular, \rhjs\
implies that one cannot distinguish
different initial states from their
long time behavior (i.e. information is
lost).

Energy eigenstates are hard to work with. It is convenient to
consider microcanonical or canonical averages of \ffj, for
example, the thermal {\it connected} Wightman functions~(see
Appendix B.1 for a precise definition of ``connected'' and the
constant $C$ below)
 \eqn\wighn{\eqalign{
 G_+ (t) & = \vev{\OO(t) \OO(0)}_{\beta}
 = {1 \ov Z} \Tr \le(e^{-\beta H} \OO (t) \OO(0) \ri) - C
 }}
or retarded functions
 \eqn\reta{
 G_R (t) = {1 \ov Z} \Tr \le(e^{-\beta H} [\OO (t), \OO(0)] \ri) \
 .
 }
We shall take the temperature $T$ to be sufficiently high so that
$E_\beta$ determined from \Diru\ lies the high energy sector.
Equation \rhjs\ implies that\foot{\rhjs\ in fact implies the
following to be true for any ensemble of states.}
 \eqn\rsn{
 G_R (t)\to 0, \qquad G_+ (t) \to 0, \qquad t \to +\infty \ .
 }
Note that $G_R (t)$
  measures the linear response of the
system to external perturbations caused by $\OO$. That $G_R (t)
\to 0$ for $t \to \infty$ implies that any small perturbation of
the system away from the thermal equilibrium eventually dies away.
In a weaker sense than \rhjs, \rsn\ can also be considered as an
indication of the emergence of an arrow of time.

In frequency space, the Fourier transform\foot{We use the same
letter to denote the Fourier transform of a function,
distinguishing them by the argument of the function.} of \wighn\
and \reta\ can be written in terms of a spectral density function
$\rho (\om)$~(see Appendix B.1 for a review)
 \eqn\Lehs{\eqalign{
 G_+ (\om) & = {1 \ov 1- e^{-\beta \om}} \rho (\om) \cr
  G_R (\om) & = -  \int_{-\infty}^\infty {d \om' \ov 2 \pi}
 {\rho (\om') \ov  \om - \om' + i \ep}
 }}
\rsn\ may be characterized by properties of the spectral density
$\rho (\om)$. For example from the Riemann-Lebesgue theorem, \rsn\
should hold if $\rho (\om)$ is an integrable function on the real
axis. Since other real-time correlation functions can be obtained
from $G_+$ (or spectral density function $\rho (\om)$) from
standard relations, for the rest of the paper, we will focus on
$G_+$ only.

For $\NN=4$ SYM at strong coupling, it
is convenient to take $\OO$ to have a
definite angular momentum $l$ on $S^3$.
\wighn\ and \reta\ can be studied by
considering a bulk field propagating in
an eternal AdS black hole geometry and
one does find the behavior \rsn\ as
first emphasized in~\refs{\maldat}. In
the bulk language, \rsn\ can be
heuristically interpreted as the fact
that any small perturbation of the
black hole geometry eventually dies
away by falling into the horizon.
Furthermore, by going to frequency
space, one finds that the Fourier
transform $G_+ (\om,l)$ has a rich
analytic structure in the complex
$\om$-plane\foot{Similar things can
also said about $G_R (\om,l)$ which can
be obtained from $G_+ (\om,l)$ using
standard relations.}, which encodes
that the bulk black hole geometry
contains a horizon and singularities.
The main features can be summarized as
follows~\refs{\guidoliu}:

\item{1.} $G_+ (\om, l)$ has a  continuous spectrum with $\om \in
(-\infty, +\infty)$. This is due to the presence of the horizon in
the bulk.

\item{2.} In the complex $\om$-plane, the only singularities of
$G_+ (\om, l)$ are poles. The decay rate for $G_+ (t)$ at large
$t$ is controlled by the imaginary part of the poles closest to
the real axis, which is of order $\beta$.

\item {3.} The presence of black hole singularities in the bulk
geometry is encoded in the behavior of $G_+ (\om,l)$ at the
imaginary infinity of the $\om$-plane\foot{See
also~\refs{\shenker} for signature of the black hole singularities
in coordinate space.}. In particular,

3a. $G_{+} (\om,l)$ decays exponentially as $\om \to \pm i
\infty$.

3b. Derivatives of $G_{+} (\om,l)$ over $l$ evaluated at $l=0$ are
divergent as $\om \to \pm i \infty$.

\ndt As emphasized in~\refs{\guidoliu}, none of the above features
 survives at finite $N$, in which case\foot{Note that even
though $\NN=4$ SYM on $S^3$ is a field
 theory, at finite $N$ the theory can be effectively considered as a theory with
 a finite number of degrees of freedom, since for any given energy $E$, there
 are only a finite number of modes below that energy. Furthermore,
 given that the number of modes with frequency ${k \ov R}$ grows with $k$ only as
 a power, it is more entropically favorable to excite modes with
 low $k$ for $E \sim O(N^2)$ and modes with $\om_\al \sim O(N)$ are almost never excited.}
$$
 G_+ (\om) = 2 \pi \sum_{m,n} e^{-\beta E_m} \rho_{mn} \delta
 (\om- E_n + E_m)
 $$
has a discrete spectrum and is a sum of delta functions supported
on the real axis. This indicates that concepts like horizon and
singularities only have an approximate meaning in a semi-classical
limit (large $N$ limit).

To understand the information loss
paradox and the resolution of black
hole singularities, we need to
understand how and why they arise in
the classical limit of a quantum
gravity. In Yang-Mills theory, this
boils down to understanding what
physics is missed in the large $N$
limit and why missing it is responsible
for the appearance of singularities and
the loss of information. With these
motivations in mind, in this paper we
are interested in understanding the
following questions

\item{1.} Can one find a qualitative argument for the emergence of
an arrow of time in the large $N$ limit?

\item{2.} Does the analytic behavior observed at strong coupling
persist to weak coupling?

\ndt which we turn to in the following sections.

\newsec{Non-thermalization in perturbation theory}

In this section we consider  \wighn\ in perturbation theory in the
planar limit. We will find that real-time correlation functions
have a discrete spectrum and  oscillatory behavior. Thus the
theory does not thermalize in the large $N$ limit.

In perturbation theory, $G_+ (t)$ can
be computed using two methods. In the
first method, one computes $G_E (\tau)$
with $0 < \tau< \beta $ in Euclidean
space using standard Feynman diagram
techniques. $G_+(t)$ can then be
obtained by taking $\tau= i t + \ep$.
An alternative way is to double the
fields and use the analogue of the
Schwinger-Keldysh contour to compute
the Feynman function $G_F (\om)$ in
frequency space~\refs{\semenoff}, from
which $G_+ (\om)$ can be obtained. In
the Euclidean-time method it is more
convenient to do the computation in
coordinate space since one does not
have to sum over discrete frequencies,
while in the
real-time method frequency space is more convenient to use. 

We look at the free theory first.

\subsec{Free theory}

To evaluate  \wighn\ in free theory, it is convenient to use the
Euclidean method.  The Euclidean correlator
 \eqn\rjs{
 G_E^{(0)} (\tau) = \vev{\OO(\tau) \OO(0)}_{0,\beta}, \qquad 0
 \leq \tau < \beta
 }
with $\OO$ of the form \fjd\ can be computed using the Wick
contraction\foot{see e.g.~\refs{\brigante} for a derivation of the
following equation and some examples of correlation functions in
free theory.}
 \eqn\fincon{\eqalign{
 & \underbrace{M_{ ij}^{\al_1} (\tau) \, M^{\al_2}_{kl} (0)}
 = {\delta_{\al_1 \al_2} \ov N} \sum_{m=-\infty}^\infty g_E^{(0)}(\tau - m \beta; \om_{\al_1})
 U^{-m}_{il} U^{m}_{kj}
 \cr
 }}
where $g_E^{(0)}$ is the propagator at zero temperature
 \eqn\Hrr{
 g_E^{(0)} ( \tau; \om) = {1 \ov 2 \om} e^{-\om |\tau|} \ .
 }
In \fincon\ $U$ is a unitary matrix
which arises due to covariant
derivatives in \onedAc\ and can be
understood as the Wilson line of $A$
wound around the $\tau$ direction. In
the evaluation of free theory
correlation functions
$\vev{\cdots}_{0,\beta}$ in \rjs, one
first preforms the Wick contractions
\fincon\ and then performs the unitary
matrix integral over $U$, which plays
the role of projecting the intermediate
states to the singlet sector. In the
large $N$ limit, the $U$ integral can
be evaluated by a saddle point
approximation. Note in particular
that~\refs{\minW}
 \eqn\fjd{
 U \to 1, \qquad T \to \infty
 }
Equation \fjd\ indicates that the
singlet condition should not play an
important role for states of
sufficiently high energy.

For definiteness, we now restrict to theories with a single
fundamental frequency $\om_0$ like $\NN=4$ SYM or \GonedAc.
Relaxing this restriction does not affect our main conclusions, as
will be commented on in various places below. Wick contractions in
\rjs\ give rise to terms of the form $e^{n \om_0 \tau}$ for some
integer $n$, while the $U$-integral computes the coefficients of
these terms. Thus \rjs\ always has the form
 \eqn\rjek{
 G_E^{(0)} (\tau) = \sum_{n=-\Delta}^{\Delta}  c_n (\beta) e^{ n \om_0 \tau}
 }
where $\Delta$ is the dimension of the operator\foot{Note that for
$\NN=4$ SYM the dimension of $M_\al$ is given by ${\om_\al \ov
\om_0}$. For other matrix quantum mechanical systems without
conformal symmetry one can use a similar definition in free
theory. For bosonic operators, $\Delta$ are integers.}.
Analytically continuing
 \rjek\ to real time, we find that
 \eqn\Gjek{
G_+^{(0)} (t) = \sum_{n=-\Delta}^{\Delta}  c_n (\beta) e^{-i n
\om_0 t} \
 }
 and
  \eqn\rjeo{
G_+^{(0)} (\om) = 2 \pi \sum_{n=-\Delta}^{\Delta}  c_n (\beta)
\delta (\om - n \om_0) \ .
 }

Thus in the large $N$ limit, the correlation function always shows
a discrete spectrum is quasi-periodic. The results are generic. If
the theory under consideration has several incommensurate
fundamental frequencies, one simply includes a sum like those
\rjek\ and \rjeo\ for each such frequency. The maximal number of
independent exponentials is $2^K$, where $K$ is the total number
of matrices in~$\OO$. This is due to that each matrix in~$\OO$ can
only connect states with a definite energy difference.

It is also instructive to obtain \Gjek\ using a different method.
By inserting a complete set of free theory energy eigenstates in
\wighn\ we find that
 \eqn\fjdk{
 G_+^{(0)} (t) = {1 \ov Z_0} \sum_{a,b} e^{-\beta \ep_a} \rho_{ab}
 e^{i (\ep_a - \ep_b) t}
 }
where $\ket{a}$  is a free theory state with energy $\ep_a$ and
$\rho_{ab} = |\vev{a|\OO(0)|b}|^2$. To understand the structure of
\fjdk\ we expand $\OO(0)$ in terms of creation and annihilation
operators associated with each $(M_\al)_{ij}$, from which we find
that

\item{A.} Due to energy conservation, $\OO$ can connect levels
whose energy differences lie between
$-\De \om_0$ and $\De \om_0$, i.e.
$\rho_{ab}$ can only be non-vanishing
for $|\ep_a - \ep_b|< \Delta \om_0$.

\item{B.} $\OO$ can only  connect states whose energy differences
are integer multiples of $\om_0$ i.e. $\rho_{ab}$ can only be
non-vanishing for $\ep_a - \ep_b = n \om_0$ with $|n| < \Delta$
integers (or half integers if $\OO$ is fermionic). Similarly, in
the cases where $\OO$ contains $K$ types of matrices of different
frequency $\om_i$ it can only connect states whose energy differs
by $\sum_{i=1}^K n_i \om_i$, where $n_i$ are integers whose
absolute values are bounded by the number of matrices of each type
appearing in $\OO$.

\ndt As a result, \fjdk\ must have the
form \Gjek. Note that the argument
based on \fjdk\ applies not only to the
thermal ensemble, but in fact to
correlation functions in any density
matrix (or pure state).

To summarize, one finds that in free theory a real-time thermal
two-point function always has a discrete spectrum and is
quasi-periodic in the large $N$ limit. This implies that once one
perturbs the theory away from thermal equilibrium, the system
never falls back and keeps oscillating. This is not surprising
since the system is free and there is no interaction to thermalize
any disturbance. Note that this is distinctly different from the
behavior \rsn\ found at strong coupling. In particular, this
implies that the bulk description of the high temperature phase in
free theory looks nothing like a black hole. Also note that the
story here is very different from that of the orbifold CFT in the
AdS$_3$/CFT$_2$ correspondence. There the mass gap in free theory
goes to zero in the large $N$ limit in the long string
sector~\refs{\maldsuss}. As a result, one finds that free theory
correlation functions in the long string sector do resemble those
from a BTZ black hole~\refs{\maldat,\vjay}.

\subsec{Perturbation theory}

In this subsection we use a simple example \GonedAc\ for
illustration. The general features discussed below apply to
generic theories in \onedAc\ including $\NN=4$ SYM.

In perturbation theory $G_E (\tau)$  can be expanded in terms of
$\lam$ as
 \eqn\rjr{
G_E (\tau) = \sum_{n=0}^\infty \lam^n G^{(n)}_E (\tau)
 }
where $G^{(0)}_E$ is the free theory result. We will be only
interested in the connected part of $G_E (\tau)$. Higher order
corrections are obtained by expanding
 $e^{- \lam \int d \tau V}$ in the path integral with $V$ given by the
 quartic term  in \GonedAc. More explicitly, a
 typical contribution to $ G^{(n)}_E (\tau)$ in \rjr\ has the form
 \eqn\ryus{
  {(-1)^{n} \ov n!}
  \int_{0}^\beta d \tau_{1} \cdots \int_{0}^\beta d
 \tau_{n} \, \vev{\OO (\tau) \OO (0) \,
  V (\tau_{1}) \cdots V
 (\tau_{n})}_{\beta,0}
 }
The free theory correlation function inside the integrals in
\ryus\ can be computed by first using Wick contraction \fincon\
and then doing the $U$ integral. The general structure of \ryus\
can be summarized as follows:

\item{1.}
 The planar diagram contribution to $G^{(n)}_E (\tau)$ scales
like $N^{0}$, while diagrams of other
topologies give higher order $1/N^2$
corrections.  The number $R_n$ of
planar diagrams grows like a power in
$n$, i.e. is bounded by $C^n$ with $C$
some finite constant~\refs{\thooft}.

\item{2.} The $\tau$-integrations are over a compact segment and
are all well defined. A typical term in \ryus\ after the
integration has the structure
 \eqn\fjdp{
g_{kj}^{(n)} (\beta) \tau^l e^{k \om_0 \tau}
 }
where $l$ and $k$ are integers. $l$ can take values from $0$ to
$n$, while $k$ from $-2n - \Delta$ to $2n + \Delta$ where $\Delta$
is the dimension of $\OO$ in free theory.

\medskip

Analytically continuing \rjr\ to Lorentzian time by taking $\tau =
it + \ep$, we find
 \eqn\dkdu{
 G_+ (t,\lam) = \sum_{n=0}^\infty \lam^n G^{(n)}_+ (t)
 }
where typical terms in $G^{(n)}_+ (t)$ have the $t$-dependence of
the form
 \eqn\rjd{
g_{kl}^{(n)} (\beta) t^l e^{i k \om_0 t}
 }
with the range of $l$ and $k$ given
after equation \fjdp. After Fourier
transforming to frequency space we find
that at each order in the perturbative
expansion $ G^{(n)}_+ (\om)$ (and thus
the spectral density function $\rho
(\om)$) consists of sums of terms of
the form
 \eqn\djs{
 g_{kl}^{(n)}   \delta^{(l)} (\om- k \om_0 )
 }
where the superscript $l$ denotes the number of derivatives.

If the theory has more than one fundamental frequencies, since the
interaction vertices are traces of a finite number of matrices
they only connect states with definite energy differences. More
and more frequencies will appear in the spectrum of a correlation
function as we go to higher and higher orders in the perturbative
expansion. The increase in the number of frequencies is
exponential in the order of the expansion but at any fixed finite
order no matter how large the spectrum of the correlation
functions is discrete.

One origin of $t^l$ terms in \rjd\ is the shifting of frequency
from the free theory value. For example, suppose the free theory
frequency is shifted to $\om = \om_0 + \lam \om_1 + \cdots$, one
would get terms of the form \rjd\ when expanding the exponential
$e^{i \om t}$ in $\lam$. One can in principle improve the
perturbation theory by resumming such contributions using Dyson's
equations. However, there appears no systematic way of doing this
for a composite operator \fjd. In Appendix B.2, we prove that
real-time correlation functions of fundamental modes $M_\al$ again
have a discrete spectrum in the improved perturbative expansion.

\newsec{Break down of Planar perturbation theory}

It is well known that at zero temperature the planar expansion of
a matrix quantum mechanics has a finite radius of convergence in
the $\lam$-plane (see e.g.~\refs{\thooft} for a recent discussion
and earlier references). If this persists at finite temperature,
properties of the theory at zero coupling or in perturbation
theory should hold at least for the coupling constant being
sufficiently small. In particular, from our discussion of last
section, one would conclude that real-time correlation functions
for generic gauge invariant operators should be quasi-periodic and
an arrow of time does not emerge at small 't Hooft coupling. In
this section, we argue that the planar perturbative expansion in
fact breaks down for real-time correlation functions and thus
perturbation theory cannot be used to understand the long-time
behavior of the system at any nonzero coupling.

From our discussion in section 3.2, we expect the Euclidean
correlation function \rjr\ should have a finite radius of
convergence for any given $\tau \in (0, \beta)$. After analytic
continuation to real time, the convergence of the expansion in
Euclidean time implies that \dkdu\ should have a finite radius
$\lam_c (t)$ of convergence for any given $t$. However, it does
not tell how $\lam_c (t)$ changes with $t$ in the limit $t \to
\infty$. In this section we argue that the radius of convergence
goes to zero in the large $t$ limit. Note that the convergence of
the perturbative expansion depends crucially on how $g^{(n)}_{kl}$
in \djs\ fall off with $n$. We will argue below that the falloff
is slow enough that perturbation theory breaks down in the long
time limit. In frequency space, one finds that $n$-th order term
in the expansion grows like $n!$\foot{Note that in frequency space
the relation between real-time and Euclidean correlation functions
is not simple, since Euclidean correlation functions are only
defined at discrete imaginary frequencies.}.

We will again use \GonedAc\ as an
illustration. The argument generalizes
immediately to generic systems
in~\onedAc. For simplicity, we will
consider the high temperature limit
\fjd\ in which we can replace $U$ in
\fincon\ by the identity matrix, e.g.
 \eqn\incon{\eqalign{
    &  \underbrace{M_{1 ij} (\tau) \, M_{1 kl} (0)}
 = {1 \ov N} \sum_{m=-\infty}^\infty g_E^{(0)} (\tau - m \beta; \om_{0})
 \delta_{il} \delta_{kj} = {1 \ov N} \delta_{il} \delta_{kj} \;
 g_E
 (\tau; \om)
 }}
where
 \eqn\hrr{
 g_E (\tau;\om) = {1 \ov 2 \om} \le(e^{-\om \tau} (1+f (\om)) +
 e^{\om
 \tau} f (\om) \ri),
 \qquad \tau \in (0, \beta)
 }
with
 \eqn\smsa{
  f(\om) = {1 \ov e^{\beta \om } - 1} \ .
 }
Note that outside the range in \hrr, $g_E (\tau)$ is periodic.

For our purpose it is enough to examine
the Wightman function for $M_1$,
 \eqn\rowp{
  D_+ (t) = {1 \ov Z (\beta)} \Tr \le(e^{-\beta H} M_1 (t) M_1 (0) \ri)
  \ .
 }
An exactly parallel argument to that of
the last section leads to the expansion
 \eqn\kdu{
 D_+ (t,\lam) = \sum_{n=0}^\infty \lam^n D^{(n)}_+ (t)
 }
where typical terms in $D^{(n)}_+ (t)$ have the $t$-dependence of
the form
 \eqn\hdd{
d_{kl}^{(n)} (\beta) t^l e^{i k \om_0 t}
 }
The convergence of series depends on how $d_{kl}^{(n)}$ fall off
with $n$. For our purpose it is enough to concentrate on the term
with the highest power $t$ in each order, i.e. the coefficients of
$t^n$ with given $k$. More explicitly, we will look at a term of
the form
  \eqn\kndu{
 D_+ (t,\lam) = D_+^{(0)} (t) \sum_{n=0}^\infty c_n \lam^n t^n + \cdots
 }
 where $D_+^{(0)}$ is the free theory expression. 

As before we will first compute \kndu\
in Euclidean time and then perform an
analytic continuation. Calculating

$c_n$ explicitly at each loop order for
all $n$ is of course impractical. Our
strategy is as follows. We will
identify a family (in fact infinite
families as we will see below) of
planar Feynman diagrams of increasing
loop order and show that their
contribution to $c_n$ falls off like a
power in $n$. Barring any unforseen
magical cancellation\foot{Note that
since we are in the high temperature
phase, in which supersymmetry is badly
broken, there is no obvious reason for
suspecting such magical
cancelations.}, this would imply that
the perturbation series \dkdu\ has
a zero radius of convergence in the $t
\to \infty$ limit. The simplest set of
diagrams which meet our purpose are
given by:

\ifig\Feyn{A family of diagrams which indicates that the
perturbation theory break down in the long time limit. Black and
red lines denote propagators of $M_1$ and $M_2$ respectively.}
{\epsfxsize=9cm \epsfbox{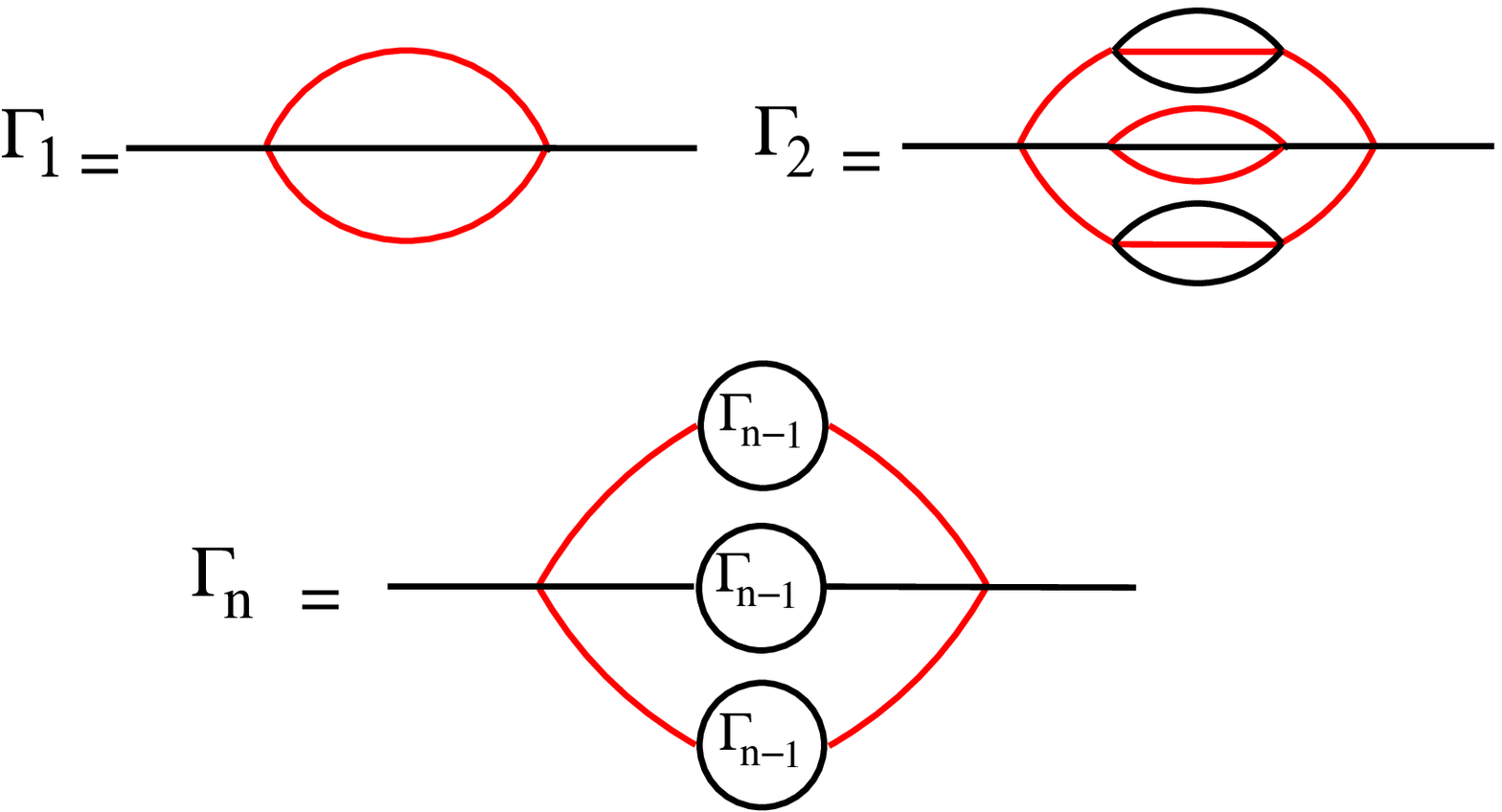}}

These graphs appear at orders $d_1=2,d_2=8,d_3=26,\cdots$ of
perturbation theory where
 \eqn\fhje{
 d_i=3d_{i-1}+2=3^i-1,  \qquad i=1,2 \cdots \ .
 }
 We denote the contribution of each
diagram by $\Ga_i (\tau)$. For our
purpose, it is not necessary to compute
the full graph. We will only need to
calculate the term in each graph with
the highest power of $\tau$, i.e. the
term proportional to $\tau^{d_i}$. Also
note that in each diagram, the symmetry
factor is exactly $1$. Let us start
with $\Ga_1$, which is given by
 \eqn\rism{
 \Ga_1 (\tau-\tau')= \lam^2 \int_0^\beta d \tau_1 d \tau_2 \, g_E (\tau - \tau_1;\om_0)
 g^3_E (\tau_1 -
 \tau_2;\om_0) g_E (\tau_2 - \tau';\om_0)
  }
Note the identity
 \eqn\fjsp{
g^3_E (\tau;\om_0) = {3\ov (2\om_0)^2}f^2(\om_0)\le(e^{\beta
\om_0} g_E(\tau;\om_0)+{f(\om_0 )\ov f(3 \om_0)} g_E(\tau;3
\om_0)\ri)
 }
Now plug \fjsp\ into \rism. It is easy to convince oneself that
the term proportional to $g_E(\tau;3\om_0)$ in \fjsp\ will not
generate a term proportional to $\tau^2$ and we will ignore it.
The contribution of the term proportional to $g_E(\tau;\om_0)$ can
be found by noting the identity\foot{If two matrices have
different frequencies in the product
$g_E(\tau;\om_0)g^2_E(\tau;\om_1)$ there is also a term
proportional to $g_E(\tau;\om_0)$ with coefficient
${1\ov(2\om_1)^2}f(\om_1)(1+f(\om_1))$ and the rest of the
analysis follows with minor changes.}
 \eqn\iduw{
  \int_0^\beta d\tau_1 d\tau_2 \;
g_E(\tau-\tau_1;\om_0)g_E(\tau_1-\tau_2;\om_0)g_E(\tau_2-\tau';\om_0)
= \ha {\p^2 \ov (\p \om_0^2)^2} g_E(\tau-\tau';\om_0)
 }
The right hand side of \iduw\ contains a piece $\ha
{(\tau-\tau')^2 \ov (2\om_0)^2} g_E(\tau-\tau')$ plus parts with
smaller powers of $\tau-\tau'$. Thus the term in \rism\
proportional to $(\tau-\tau')^2$ is given by
 \eqn\rijs{
 \Ga_1 (\tau-\tau')= {\al \lam^2 \ov 2 }  (\tau
 -\tau')^2 g_E (\tau-\tau') + \cdots
 }
where
 \eqn\rimn{
\al = {3 f (1+f) \ov (2 \om_0)^4}, \qquad f = f(\om_0)
 }

The term proportional to $\tau^{d_i}$ for higher order diagrams
$\Ga_i (\tau)$ can now  be obtained by iterating the above
procedure. A useful identity is
 \eqn\jwpw{\eqalign{
 & \int_0^\beta d\tau_1 d\tau_2
\; g_E(\tau-\tau_1;\om_0)g_E(\tau_1-\tau_2;\om_0)\, (\tau_1 -
\tau_2)^n \, g_E(\tau_2-\tau';\om_0) \cr
 & = {(\tau-\tau')^{n+2}\ov(2\om_0)^2}{1\ov (n+2) (n+1)}
 g_E(\tau-\tau';\om_0)+ \cdots
 }}
where we kept only the term with the highest power of
$\tau-\tau'$, as lower power terms will not contribute to the
terms in which we are interested. We find that the term
proportional to $\tau^{d_i}$ in $\Ga_i (\tau)$ is given by
 \eqn\fjjs{
 \Ga_i (\tau) = F_i  \lam^{d_i}  \tau^{d_i} g_E (\tau;\om_0)+ \cdots
 }
 where $F_i$ satisfy the recursive relation
 \eqn\dses{
 F_{i+1} = F_i^3  {\al \ov d_{i+1} (d_{i+1}-1)} \ .
  }
Thus $F_i$ can be written as
 \eqn\fkks{
 F_i =  \al^{d_i \ov 2} \Lam_i
 }
with
 \eqn\nsoe{
 \Lam_i =
 \prod_{k=0}^{i-1} \le({1 \ov d_{i-k} (d_{i-k}-1) } \ri)^{3^k}
 }
$\Lam_i$ in the large $i$ limit can be easily estimated and we
find
 $$
 \Lam_i \approx e^{-{3 \ov 2} d_i} , \qquad i \gg 1 \ .
 $$

Summing all our diagrams together and analytically continuing to
Lorentzian time with $\tau=i t + \ep$, we find that\foot{Since we
are only interested in the asymptotic behavior of the sum for
large $i$ we have replaced $F_i$ by its asymptotic value.}
 \eqn\dkwi{
 \sum_i \Ga_i (t) \approx D_+^{(0)} (t)
 \sum_{i=1}^\infty  (-1)^i \le({\lam t \ov h_c}\ri)^{d_i} + \cdots
 }
with $h_c$ given by
 \eqn\dkkr{
 h_c = {e^{3 \ov 2} \ov \sqrt{\al}} ={e^{3 \ov 2} (2 \om_0)^2 \ov  \sqrt{3 f
 (1+f)}}
 }
Equation \dkwi\ implies that the radius
of convergence in $\lam$ is given by
 \eqn\hkro{
 \lam_c (t) \sim {1 \ov t}
 }
which goes to zero as $t \to \infty$.

It is also instructive to repeat the
computation of \Feyn\ in frequency
space using the real-time method. The
calculation is straightforward and we
will only summarize the result. One
finds that the contribution of $\Ga_i$
to the Feynman function $D_F (\om)$
grows like $d_i !$. Thus one expects
that the perturbative expansion in
frequency space is not well defined for
any frequency.  Note that the
non-analyticity in frequency space can
be expected since in going to frequency
space one has to integrate the full
real time-axis and the Fourier
transform is sensitive to the long time
behavior. Also note that the $n!$
factorial behavior in perturbation
theory often implies an essential
singularity at $\lam =0$ (see also
below).

We conclude this section with some remarks:

\item{1.} In the zero-temperature limit $h_c \to \infty$ and the
set of terms in \dkwi\ all go to zero.

\item{2.} To simplify our discussion, we have only considered
diagrams in \Feyn. There are in fact many other diagrams of
similar type contributing at other orders in $\lam$. For example,
by including those in fig.2, one can get contributions for all
even orders in $\lam$ rather than only \fhje. The qualitative
conclusion we reached above is not affected by including
them\foot{There are also potentially an infinite number of other
sets of diagrams which can lead to the behavior \hkro, e.g. one
can replace $\Ga_1$ by any diagram whose highest power in $t$ is
the same as the order of perturbation and then iterates.}.
 \ifig\FeynN{By including the diagrams on the left with all possible $i,j,k \geq 0$ we can
 get contribution at every even order of $\lam$ instead of \fhje. $\Ga_0$ denotes a single
 propagator. Diagrams on the right can also contribute to the odd orders
 if \GonedAc\ contains additional interactions of the form $\tr
 A^2 B^2$.
 }
{\epsfxsize=10cm \epsfbox{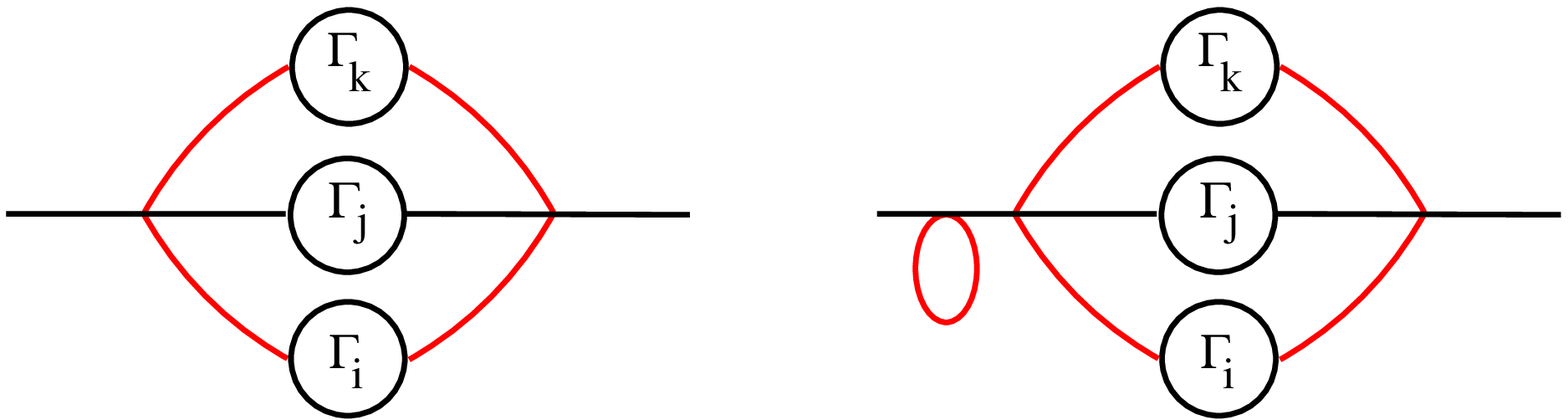}}

\item{3.} By taking in consideration the diagrams on the left of \FeynN\ the sum in \dkwi\ is extended
\foot{the value of $h_c$ also changes}
to all even powers of $\lambda t$ and
is oscillating therefore the
singularities in $\lam t$ should lie on
the imaginary axis. Let us suppose that
for a given $\lam$, $D_+ (t)$ has a
singularity in $t$ at $q_c/\lam$ with
$q_c$ lying in the upper half
plane\foot{Note that $q_c^*/\lam$ must
also be a singularity of $D_+ (t)$.}.
Now Fourier transforming $D_+ (t)$ we
find that
 \eqn\uoek{
 D_+ (\om) = \int_{-\infty}^\infty dt \, e^{i \om t} \, D_+ (t)
 }
The presence of $q_c/\lam$ and $q_c^*/\lam$ implies $D_+(\om)$
contains a term of the form for $\om >0$
 \eqn\euns{
 D_+ (\om) \sim e^{i \om {q_c \ov \lam}}
 }
Thus $D_+ (\om)$ contains an essential singularity at $\lam =0$.

\item{4.} The $n!$ behavior in perturbative expansion in frequency
space (say in the computation of $D_F (\om)$) arises from a {\it
single} class of Feynman diagrams. This is reminiscent of
renormalons in field theories~\refs{\gross,\lautrup,\tHooftRE}. In
particular, when Borel resumming the divergent series, depending
on whether $\om$ is greater or smaller than $\om_0$, the
singularities on the Borel plane can appear on the positive or
negative real axis\foot{Since we only have contributions to even
order in $\lam$, we cannot make a conclusion from our discussion
so far.}, also reminiscent of the IR and UV renormalons.

\item{5.} Note that in the limit $T \to \infty$, $h_c$ in \dkkr\
scales with $T$ as $h_c \sim {\om_0^3 \ov T}$, i.e.
 \eqn\ens{
 \lam_c (t) \sim {\om_0^3 \ov t T}
 }
For fixed $\lam$, we expect a singularity for $D_+ (t)$ at
 \eqn\ejs{
 t \sim {\om_0^3 \ov \lam T}
 }
Note that the right hand side of \ejs\ is reminiscent of the
magnetic mass scale for a Yang-Mills theory (see
e.g.~\refs{\GrossBR}). However, in our matrix quantum mechanics,
there is no infrared divergence and it is not clear whether there
is a connection.

\item{6.} The discussion can be straightforwardly applied to a
generic theory in \onedAc\ with cubic and quartic couplings.
 In fact
the argument also applies to a single anharmonic oscillator at
finite temperature, even though in that case one does not expect
the perturbative expansion to converge anyway\foot{See Appendix D
for further elaborations on the example of a  single anharmonic
oscillator and a discussion on the differences between the single
anharmonic oscillator and the matrix systems under consideration.
}. Similarly, the argument also applies to a single-matrix quantum
mechanics if one does not impose the singlet condition. When
imposing the singlet condition, the matrix $U$ in equation
\fincon\ cannot be set to $1$ and our argument does not apply.
Similarly our argument does not apply to \onedAc\ in the low
energy sector, in which $U$ is always important. Indeed using the
results of~\refs{\brigante,\furuuch}, one can show that to leading
order in the large $N$ limit, correlation functions at finite
temperature can be written in terms of those at zero temperature
and we do expect that the perturbation theory has a finite radius
of convergence.

\item{7.} Our argument indicates that perturbation theory breaks
down in the long time limit for a generic theory in \onedAc.
However, for any specific theory (say $\NN=4$ SYM theory) we
cannot rule out magical cancelations which could in principle
make the coefficients of $n$-th order term much smaller than
indicated by the diagrams we find. If magical cancelations do
occur in some theory, that would also be extremely interesting
since it indicates some hitherto unknown hidden
structure\foot{Since we are working at a finite temperature,
supersymmetry alone should not be sufficient for the
cancelations.}.

\newsec{Physical explanation for the breakdown of planar
expansion}

In this section we give an alternative argument for the breakdown
of perturbation theory in the long time limit, which complements
that of last section. The discussion below should apply to a
generic theory in \onedAc. For definiteness we use $\NN=4$ SYM as
an illustration example.

We first set up some notations. We write the full Hamiltonian as
 \eqn\hsja{
 H = H_0 + V (\lam)
 }
 with $H_0$ the Hamiltonian of the free theory and $V$ the interaction.
 We denote a free theory energy
eigenstate by $\ket{a}$ with energy
$\ep_a$. $\ket{0}$ is the (unique) free
theory vacuum. The energy eigenstates
of the interacting theory $H$ are
denoted by $\ket{i}$ with energy $E_i$.
$\ket{\Om}$ is the interacting theory
vacuum. We can expand
 \eqn\rdj{
 \ket{i} = \sum_a c_{ia} \ket{a}
 }
 with $c_{ia}$ satisfying
 \eqn\nska{
 \sum_a |c_{ia}|^2 = \sum_i |c_{ia}|^2 =1 \ .
 }

\ifig\specT{The energy spectrum of free $\NN=4$ SYM on $S^3$ is
quantized. Typical degeneracy for an energy level $\ep \sim O(1)$
is of order $O(1)$.  Typical degeneracy for a level of energy $\ep
\sim O(N^2)$ is of order $e^{O(N^2})$.} {\epsfxsize=3cm
\epsfbox{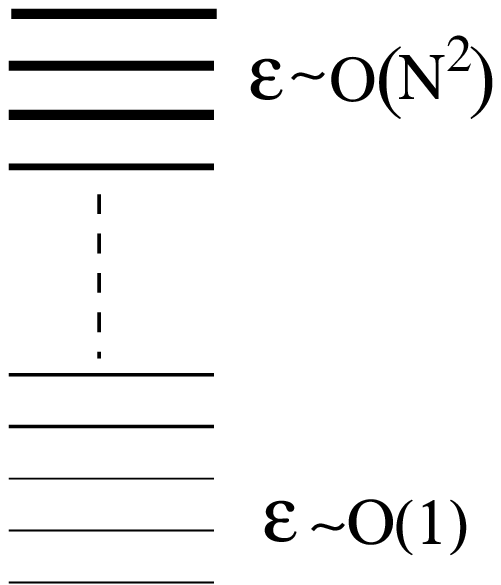}}

We first recall some relevant features
of the free theory energy spectrum of
$\NN=4$ SYM on $S^3$. Since $\om_\al$
in \onedAc\ are all integer or
half-integer multiples of $\om_0 = {1
\ov R}$, the free theory energy
spectrum is quantized in units $\ha
\om_0$. Typical energy levels are
degenerate. The degeneracy is of $O(1)$
in the low energy sector and of order
$e^{O(N^2)}$ in the high energy sector.
The exponentially large degeneracy in
the high energy sector can be seen as
follows. From \Diru\ the density of
states $\Om_0 (\ep)$ in the high energy
sector is of order $e^{O(N^2)}$. Since
the energy levels are equally spaced
with spacings order $O(1)$, it must be
that typical energy levels have a
degeneracy of order $e^{O(N^2)}$.
Alternatively, the number of ways to
construct a state of energy of order
$O(N^2)$ from $O(N^2)$ oscillators of
frequency of $O(1)$ is clearly
exponentially large in $N^2$.

When a theory contains $n
>1$ incommensurate fundamental frequencies, the free theory spectrum at energies of
order $O(N^2)$ will have level spacings of order $O(N^{-2(n-1)})$.
Since the density of states are of order $e^{O(N^2)}$, the
degeneracy of a typical state is again of order $e^{O(N^2)}$ as in
the case of one fundamental frequency. That the level spacings go
to zero as a negative power in $N$ does not change our conclusion
of previous sections regarding the thermalization in free theory
or at any finite order in perturbation theory, since as emphasized
there a small operator can only connect states whose energy
differences are of order $O(N^0)$. Therefore such small level
spacings cannot be accessed dynamically. The restriction above to
a finite number of incommensurate fundamental frequencies is not
essential. The conclusion applies to any theories in which the
number of fundamental frequencies increases only as a power of the
frequency. For these theories, the thermal ensemble is dominated
by states built from oscillators whose frequencies are smaller
than or of order of $\beta^{-1}$. Thus the effective number of
fundamental frequencies is finite.

Now let us turn on the interaction $V(\lam)$ \higT\ with a tiny
but nonzero $\lam$. We will focus on the high energy sector. Given
that free theory energy levels are highly degenerate, one would
like to apply degenerate perturbation, say to diagonalize $V$ in a
degenerate subspace of energy $E \approx \mu N^2$ and of dimension
$e^{O(N^2)}$. For this purpose we need to choose a basis in the
degenerate subspace. This is a rather complicated question, due to
difficulties in imposing singlet conditions\foot{The trace
relations are important for states of such energies.}. However,
when $\mu$ is sufficiently large we expect the singlet condition
not to play an important role\foot{As remarked earlier, in the
high temperature limit the saddle point for $U$ (in \fincon)
approaches the identity matrix.}. So to simplify our discussion we
will ignore the singlet condition below. A convenient orthonormal
basis of energy eigenstates for $H_0$ are then monomials of
various oscillators (appropriately normalized), i.e.
 \eqn\monom{
 \prod_{\al} \prod_{i,j=1}^N \le(M_{\al ij}^\dagger
 \ri)^{n_{\al ij}}
 \ket{0} \ .
 }

 In the basis \monom, if the full theory is not integrable, $V$ can
be effectively treated as an
(extremely) sparse random
matrix\foot{Here we restrict $V$ to a
single energy level. When including all
energy levels $V$ is banded and sparse.
The banded structure is due to energy
conservation.}. The sparseness is due
to that each term in $V$ can connect a
given monomial state to at most $N^k$
other states, where $k$ is an $O(1)$
number\foot{This is because each term
in $V$ is a monomial of a few
matrices.}. Randomness has to do with
the large dimension of the subspace and
to the fact that there is no preferred
ordering for the states within the same
subspace. Diagonalizing $V$, we thus
expect, from general features of a
sparse random matrix (see Appendix C
for a summary),

 \item{1.} The
degeneracy of the free theory will
generically be broken\foot{ For
Yang-Mills theories on $S^3$, there are
remaining degeneracies associated with
the isometry group $SO(4)$ of $S^3$.
Except when one considers the sectors
with very large angular momenta on
$S^3$, typical representations of
$SO(4)$ are rather small and should not
affect our general argument.}.

\item{2.} A number of states of order $e^{O( N^2)}$ will mix under
the perturbation.

\item{3.} The typical level spacing between energy levels should
be proportional to the inverse of the density of states and is
thus exponentially small, of order $e^{-O(N^2)}$. Note that this
is exponentially smaller than the level spacings in a free theory
with a finite number of incommensurate fundamental frequencies.

The story is in fact a little more intricate. We expect the
degenerate perturbation to be a good guide if the spread of energy
eigenvalues after diagonalizing $V$ in a subspace is smaller than
the spacings between nearby energy levels. The spread $\Ga$ of
eigenvalues of $V$ can be estimated by (see Appendix C)
 \eqn\hee{
 \Ga^2 \sim \Ga^2_a = \sum_{a \neq b} |\vev{a|V|b}|^2 \sim O(N^2)
 }
for any nonzero $\lam$, where the sum restricts to a degenerate
subspace. Note that~\hee\ only depends on that $V$ is a single
trace operator and does not depend on the specific structure of
it. That $\Ga \sim O(N)$ implies that it is not sufficient to
diagonalize $V$ within a degenerate subspace. It appears more
appropriate to diagonalize\foot{This statement is of course only
heuristic since there is no sharp criterion to decide what should
be the precise size of the subspace. However, we expect the $N$
scaling should be robust.} it in a subspace with energy spread of
order $O(N)$. Thus in addition, we expect that:

\item{4.} an interacting theory eigenstate $\ket{i}$ is strongly
coupled to free theory states $\ket{a}$ within an energy shell of
order $O(N)$.

\ndt This statement will be justified
in the next section from a somewhat
different perspective. That $\Ga \sim
O(N)$ for any nonzero $\lam$ in the 't
Hooft limit indicates a tiny $\lam$ may
not really be considered as a small
perturbation after all.

Various features discussed above when turning on a small $\lam$
are clearly non-perturbative in nature. However, it may be hard to
probe them directly using Euclidean space observables like
partition functions and Euclidean correlation functions. These
observables probe only average behaviors within an energy
difference range of order $O(T)$ or larger and thus may not be
sensitive to the changes in level spacings at smaller
scales\foot{Of course if one is able to compute Euclidean
observables exactly, one should be able to extract all the
interesting physics. After all, real-time observables can be
obtained from Euclidean ones by analytic continuation. It is just
often the case that real-time physics is encoded in a very subtle
way in Euclidean observables.}. In contrast, real-time correlation
functions are much more sensitive. For example, consider the
Lehmann spectral decomposition of $G_+ (t)$, i.e.
   \eqn\fso{\eqalign{
 G_+ (t) & = {1 \ov Z} \Tr \le(e^{-\beta H} \OO (t) \OO(0) \ri)
 \cr
  & = {1 \ov Z} \sum_{i,j} e^{-\beta E_i + i (E_i- E_j) t}
  |\vev{i|\OO(0)|j}|^2
 }}
where we have inserted complete sets of energy eigenstates
 $\ket{i}$ of the interacting theory. From \fso, it is clear that
 $G_+ (t)$ can in principle probe any
small energy differences, provided one takes $t$ to be large
enough. This explains the breakdown of perturbation theory in the
long time limit observed in $G_+(t)$. At large $N$, the $\lam \to
0$ and $t \to \infty$ limits do not commute.


In this and the last sections we have presented two lines of largely
independent arguments that suggest that planar perturbation
theory breaks down in the long time limit.
 The first argument (last section) is based on an honest Feynman diagram
calculation, which establishes the breakdown of perturbation
theory, but does not tell us directly whether the spectral
functions are continuous or not. The second argument (this
section), based on the energy spectrum, is more heuristic, but
implies that the spectral functions are continuous in the large
$N$ limit. There are reasons to believe that the two arguments
should be closely related. As stressed earlier, the breakdown of
perturbation theory from the class of Feynman diagrams considered
in the last section only happens at a sufficiently high temperatures,
at which the thermal ensemble is dominated by states of energy of
order $O(N^2)$ and the energy spectrum becomes
quasi-continuous\foot{Applied to the theory at zero temperature or
at a temperature below the deconfinement temperature, the class of
diagrams gives a convergent contribution. At such a temperature,
the thermal ensemble is dominated by states of energy of order
$O(N^0)$, which are not quasi-continuous when turning on
interactions.}. Nevertheless, a precise relation between the two
arguments is not clear at this point, as they use very different
languages (one Feynman diagrams and the other energy levels).
It would be very desirable to find a direct connection and to have
an understanding of the time scale \ejs\ from the
point of view of the energy levels.

\newsec{A statistical approach}

The argument of section 4 shows that the planar perturbation
theory breaks down in the large time limit, but it does not tell
us what the long time behavior is. Non-perturbative tools are
needed to understand the long time behavior of real time
correlation functions in the large $N$ limit. Here we develop a
statistical approach, taking advantage of the extremely large
density of states in the high energy sector. In this section we
outline the main idea and the results, leaving detailed
calculations to various appendices. The statistical approach
enables us to derive some qualitative features satisfied by the
Wightman function for a generic operator at finite temperature,
including that it has a continuous spectral density function and
should decay to zero in the long time limit. The features we find
here are also shared by the Wightman function at strong coupling
found from supergravity analysis.

Our starting point is the Lehmann spectral decomposition of $G_+
(t)$ \fso,
   \eqn\fjso{\eqalign{
 G_+ (t)
  & = {1 \ov Z} \sum_{i,j} e^{-\beta E_i + i (E_i- E_j) t}
 \rho_{ij}
 }}
where
 \eqn\jdn{
  \rho_{ij} = |\vev{i|\OO(0)|j}|^2 = |\OO_{ij}|^2
 }
In momentum space
 \eqn\fkpo{
G_+(\om) = {1 \ov Z} \sum_{i,j} e^{-\beta E_i} \delta(\om + E_i-
E_j)  \rho_{ij} \ .
 }
Matrix elements $\OO_{ij}$ can in turn
be expressed in terms of those of free
theory using ($c_{ia}$ was introduced
in \rdj)
 \eqn\kdjs{
 \OO_{ij} = \vev{i|\OO(0)|j} = \sum_{a,b} c_{ia}^* c_{jb}  \vev{a|\OO|b}
  = \sum_{a,b} c_{ia}^* c_{jb} \OO_{ab}
 }
where we have inserted complete sets of free theory states and
$\OO_{ab} =\vev{a|\OO(0)|b}$.

Since for sufficiently high
temperature, the sums in \fkpo\ and
\kdjs\ are peaked at an energy with an
extremely large density of states, one
should be able to obtain the
qualitative behavior of $\rho_{ij}$ and
$G_+ (\om)$ from statistical properties
of $\OO_{ab}$ and $c_{ia}$. As
discussed in the last section, in the
interacting theory, we expect typical
level spacings scale with $N$ like
$e^{- O(N^2)}$. In the large $N$ limit,
$E_i$ can be considered as taking
continuous values. Note that this by
itself does not imply that $G_+ (\om)$
has a continuous spectral
decomposition, since it is possible
that $\rho_{ij}$ only has support for
states with finite energy differences.
We argue below that $\rho_{ij}$ has
nonzero support between states with any
$E_i - E_j \in (-\infty, \infty)$,
which is independent of $N$, and thus
$G_+ (\om)$ does have a continuous
spectrum.

Let us first look at the statistical behavior of $c_{ia}$. For
this purpose, consider the following density functions
 \eqn\fjs{
 \rho_a (E) = \sum_i |c_{ia}|^2 \delta (E-E_i)
  }
  \eqn\fksm{
  \chi_i ( \ep) = \sum_{a} |c_{ia}|^2 \delta
  (\ep - \ep_a)
  }
$\rho_{a} (E)$, first introduced by Wigner~\refs{\wigner}, is also
called the local spectral density function or strength function in
the literature\foot{These density functions have been frequently
used in quantum chaos literature, see e.g.~\refs{\casati}}. Using
normalization properties of $c_{ia}$, one finds that
 \eqn\ioep{
\int dE \, \rho_{a} (E) = 1, \qquad \int d \ep \, \chi_i ( \ep) =
1
 }
$\rho_{a} (E)$ can be considered as the distribution of
interacting theory eigenstates of energy $E$  coupling to a free
theory state $\ket{a}$. Similarly, $\chi_i (\ep)$ gives the
distribution of free theory states of energy $\ep$  coupling to an
exact eigenstate $\ket{i}$. The mean and the variances of the two
distributions are given by
 \eqn\ekDk{\eqalign{
 & \bar E_a = \int dE \, E \, \rho_{a} (E) = \vev{a|H|a} 
  }}
 \eqn\fkDk{\eqalign{
 & \sig_a = \Ga_a^2 = \int dE \, (E - \bar E_a )^2 \, \rho_{a} (E) =
\sum_{b \neq a} |\vev{a|V|b}|^2 \cr
 }}
\eqn\egKk{\eqalign{
 & \bar \ep_i = \int d \ep \, \ep \, \chi_i (\ep) = E_i -
 \vev{i|V|i} \cr
 }}
 \eqn\fgKk{\eqalign{
 & \Sig_i = \Delta_i^2 = \int d \ep \, (\ep - \ep_i)^2 \, \chi (E_i, \ep)
= \sum_{j \neq i} |\vev{i|V|j}|^2 \cr
 }}
$\bar E_a$ and $\Ga_a$ give the center and the spread of
interacting theory energy eigenstates coupling to a free state
$\ket{a}$. Similarly, $\bar \ep_i$ and $\Delta_i$ give the center
and the spread of free theory states coupling to an interacting
theory energy eigenstate $\ket{i}$.  $\Ga_a$ can be considered as
a measure of correlation among energy levels of the interacting
theory (since states whose energies differ by $\Ga_a$ could couple
to the same free theory state and are thus correlated). $\De_i$
characterizes the range of free theory states which are mixed by
perturbation. Note that the heuristic discussion after equation
\hee\ implies that $\Delta_i \sim O(N)$, which we will confirm
below using a different method.

Individual energy eigenstates are
rather hard to work with. We will
consider microcanonical averages of
\fjs\ and \fksm. After all, for \fjso\
and \fkpo\ we only need the behavior of
$\rho_{ij}$ averaged over states of
similar energies. We will denote the
average\foot{More explicitly, the
average can be written as
 \eqn\fAsm{
  \chi_E ( \ep) = {1 \ov \Om (E)} \sum_{E_i \in (E -\delta, E +\delta)} \chi_i (\ep)
  }
where $\delta$ is small enough that $\Om (E)$ does not vary
significantly in the range $(E-\delta, E+\delta)$.} of $\chi_i
(\ep)$ over interacting theory states of energy $E$ by $\chi_E
(\ep)$ and similarly the average of $\rho_a (E)$ over free theory
states $\ket{a}$ of similar energy $\ep$ by $\rho_\ep (E)$. Since
the averages involve a huge number of states and the large $N$
limit is like a thermodynamic limit in the high energy sector, we
will assume that $\chi_E (\ep)$ is a smooth {\it slow}
function\foot{Note that a function $f(E)$ is considered a slow
function if it can be written in a form $f(E) = N^a g(E/N^2)$,
where $g(x)$ is a function independent of $N$.} of $E$, i.e. it
depends on $E$ only through $E/N^2$. Similarly $\rho_{\ep} (E)$ is
assumed to depend on $\ep$ only through $\ep/N^2$. The center and
variance of $\chi_E (\ep)$ and $\rho_\ep (E)$ will be denoted by
$\bar \ep (E)$, $\Sig (E) = \Delta^2(E)$, $\bar E (\ep)$, and
$\sig (\ep) = \Ga^2 (\ep)$ respectively\foot{which can also be
obtained by the average of various quantities \ekDk-\fgKk\ to
leading order in large $N$.}. These quantities should also be slow
functions of $E$ or $\ep$ as they inherit the property from
$\chi_E (\ep)$ and $\rho_\ep (E)$. In the Appendix E we estimate
these quantities and find that
 \eqn\dbrr{\eqalign{
 \bar \ep (E) & =  N^2 g(\lam, E/N^2) \cr
 \Sig (E) & = N^2 h (\lam, E/N^2) \cr
 \bar E (\ep) & =  N^2 \tilde g (\lam, \ep/N^2) \cr
 \sig (\ep) & = N^2 \tilde h (\lam, \ep/N^2)
 }}
We emphasize that the large $N$
scalings above only depend that $V$ is
given by $N$ times single trace
operators. Given that the underlying
theory is not integrable and the
extremely large number of states, we
will thus approximate $c_{ia}$ for
fixed $i$ as a random unit vector which
centers at $\bar \ep_i$ with a spread
of order $\Delta_i \sim O(N)$.

Now we turn to the statistical
properties of $\OO_{ab}$. Our earlier
discussion for $V$ in the free state
basis \monom\ can be carried over to
any operator $\OO$ of dimension $O(1)$.
Thus $\OO_{ab}$ can be considered as an
sparse banded random matrix. The matrix
is banded since from energy
conservation $\OO$ can only connect
states whose energy difference is
smaller than the dimension of $\OO$.
Note that even though $\OO_{ab}$ is
sparse, for each row (or column), the
number of nonzero entries grows with
$N$ as a power.

To summarize, we will assume the following statistical properties
for $c_{ia}$ and $\OO_{ab}$:

\item{1.} For a given $i$, $c_{ia}$ is a random unit vector with
support inside an energy shell of width
$O(N)$. In particular, the $c_{ia}$
satisfy the same distribution for
$\ket{a}$ of the same energy.

\item{2.} $\OO_{ab}$ is banded sparse random matrix, with the
number of nonzero entries growing with $N$ as a power.

\ndt Now consider any two states
$\ket{i}$ and $\ket{j}$, with energies
$E_i$ and $E_j$ respectively, for which
$\om= E_i - E_j \sim O(1)$.  One finds
that $\bar \ep_i - \bar \ep_j \sim
O(1)$ and the energy shells of the two
states overlap significantly. Given
that the number of nonzero entries in a
row or column of $\OO_{ab}$ grows with
$N$ as a power and that each element of
$c_{ia}$ satisfies the same
distribution, one concludes from \kdjs\
that $\OO_{ij}$ should have support for
any $\om = E_i - E_j \sim O(1)$ and
$G_+ (\om)$ has a continuous spectrum
for $\om \in (-\infty, + \infty)$.
Note that the fact that $\Delta \sim
O(N)$ is crucial for having a
continuous spectrum  $\om \in (-\infty,
+ \infty)$. Suppose $\Delta \sim O(1)$,
the spectrum cannot extend to $\pm
\infty$ due to energy conservation.

One can further work out more detailed
properties of $\rho_{ij}$. Leaving the
detailed calculation in various
appendices, we find that (after
averaging $\rho_{ij}$ over states of
similar energies)
 \eqn\ehr{
 \rho_{E_1 E_2}  = {1 \ov \Om (E)} A(\om; E) = e^{- S(E)} A
 (\om; E)
 }
where $\Om (E)$ and $S(E)= \log \Om (E)$ are the density of states
and entropy of the interacting theory respectively and
$$
 E = {E_1 + E_2 \ov 2}, \qquad \om = E_1 - E_2 \ .
 $$
 Equation \ehr\ is derived in Appendix G along
with properties of
 $A (\om; E)$ stated below. Some useful formulas used in the
 derivation are collected in Appendix F.
 $A (\om; E)$ can be expressed in terms of an integral of $\chi_E (\ep)$ and $\bar \ep
 (E)$~(see equations (G.3) and (G.8))
and satisfies the following properties:

 \item{1.} $A (\om;E)$ is an even function of $\om$, i.e.
 \eqn\rjd{
 A (-\om; E) = A (\om; E)
 }

 \item{2.} As $\om \to \infty$
 \eqn\ejf{
 A (\om; E) \propto e^{-\ha \beta (E) |\om|}, \qquad \beta (E) =
 {\p S (E) \ov \p E}
 }

\item{3.} $A (\om, E)$ is integrable along the real axis and can
at most have integrable singularities of the form
 \eqn\ejjsP{
 A(\om; E) \propto {1 \ov |\om- \om_s|^{\al_s}}, \qquad \al_s < 1
 \ .
 }

\item{4.} $A_E (\om)$ depends on $E$ only through $E/N^2$, i.e. it
can be written as
 \eqn\rke{
 A (\om; E) = A(\om; \mu), \qquad \mu = {E \ov N^2}
 }
 and $A$ is a function independent of $N$.

\ndt Note that property 2 implies that in the large $N$ limit,
$\rho_{E_1 E_2} \sim 0$ for $E_1-E_2  \sim N^a$ with $a>0$.

The expression for $G_+ (\om)$ in
momentum space can now be obtained by
plugging \ehr\ into \fkpo\ and using a
saddle point approximation. We find
that
 \eqn\arnd{\eqalign{
 G_+ (\om) & = {1 \ov Z} \int dE \, e^{-\beta E} \, e^{S(E) + S(E+\om)}
  e^{- S(E + \om/2)} A (\om, E/N^2) \cr
 & = {1 \ov Z} \int dE \, e^{- \beta E + S(E)} \, \le[e^{S(E+\om)
 - S(E + \om/2)} A (\om, E/N^2) \ri] \cr
 & = e^{{\beta \om \ov 2}} A (\om, \mu_\beta)
 }}
 where
 \eqn\rhr{
 \mu_\beta = {E_{\beta} \ov N^2}, \qquad {\p S(E) \ov \p E}\biggr|_{E_\beta} =
 \beta \ .
 }
Note that since in the large $N$ limit, $E$ can be treated as
continuous and $\rho_{ij}$ has support for any energy difference,
it is appropriate to approximate the sum in \fkpo\ by an integral.
Also from the second line to the third line we have used that the
quantity inside the bracket depends on $E$ slowly and performed a
saddle point approximation. We conclude this section with some
remarks:

\item{1.} $G_+ (\om)$ has a continuous spectrum with $\om \in
(-\infty, + \infty)$ in the large $N$ limit (note that $\om$ does
not scale with $N$).

\item{2.} Since $A(\om, \mu)$ can at most have integrable
singularities of the form \ejjsP\ on the real axis, after a
Fourier transform to coordinate space, $G_+(t)$ must go to zero in
the limit $t \to \infty$. If $A (\om; \mu)$ is a smooth function
on the real axis, then $G_+(t)$ must decay exponentially with
time.

\item{3.}  Considering the last line of \arnd\ as a definition for
$A (\om; \mu)$, for $\NN=4$ SYM on $S^3$ at strong coupling, the
corresponding $A(\om; \mu)$ can be found by solving the Laplace
equation for a scalar field in an AdS black hole geometry and be
expressed in terms of boundary values of renormalizable wave
functions for the scalar field~\refs{\guidoliu}. In particular, $A
(\om;\mu)$ found at strong coupling satisfy all the properties
\rjd--\rke\ (it is a smooth function on the real axis).

\item{4.} It should be possible to obtain an explicit expression
for $A(\om; \mu)$ (and thus $G_+ (\om)$) using the expressions
found in the appendices (e.g. equation (G.3)) if one can find the
density functions \fjs\ and \fksm\ for a sparse banded random
matrix with varying density of states. While those for constant
density of states have been discussed in the literature~(see
e.g.~\refs{\casati}), not much appears to be known for the
non-constant density of states~\refs{\adam}.

\newsec{Discussions}

In this paper we first showed that in perturbation theory,
real-time correlation functions in the high temperature phase
of \onedAc\ have a discrete spectrum and the system does not
thermalize when perturbed away from thermal equilibrium. We then
argued that the perturbative expansions for real-time correlation
functions break down in the long time limit. The breakdown of
perturbation theory indicates that at large $N$ the $\lam \to 0$
and $t \to \infty$ limits do not commute. The reason for the
breakdown is that a wide energy range (of order $O(N)$) of
degenerate free theory energy eigenstates mix under the
interaction. The level spacings in the energy spectrum of $O(1)$
in the free theory become $e^{-O(N^2)}$. As a result, real-time
correlation functions develop a continuous spectrum for any
nonzero $\lam$. The continuous spectrum was argued from a
statistical approach developed in section 6, where we also  show
that real-time correlation functions should decay to zero as $t
\to \infty$ and the system becomes time irreversible.

We should emphasize that our arguments in this paper are
qualitative in nature and far from foolproof.  For example,
instead of being a random vector, $c_{ia}$ could have some
structure (e.g. being very sparse) within the range of its spread,
in which case our statistical argument will not be valid.

It is also important to emphasize our results only apply to the
high energy sector and in the low energy sector (or in the low
temperature), there is no indication of breakdown of the planar
expansion. In particular the results we describe here are not
inconsistent with that the sector near the vacuum might be
integrable in the large $N$ limit~\refs{\minahan}. In fact the
results of~\refs{\brigante,\furuuch} are consistent with that the
interpolation between the free theory and strong coupling  may be
smooth in the low temperature phase.

Our results indicate that there is a large $N$ ``phase
transition''
at $\lam =0$, i.e. physical observables undergo qualitative
changes in the limit $\lam \to 0$. The ``phase transition'' we
find here is somewhat unusual, since it is not manifest in the
Euclidean quantities like the partition function. The partition
function appears to be smooth in the $\lam \to 0$ limit. The
``phase transition'' is in real-time correlation functions and
their Fourier transforms. Real-time correlation functions decay to
zero at large time at any finite $\lam$, while oscillatory for
$\lam =0$. In frequency space there is an essential singularity at
$\lam =0$.

It would be interesting to understand
whether one can continue the physics at
small $\lam$ to large $\lam$. If there
is no further large $N$ ``phase
transition'' in $\lam$, we expect that
the analytic structure of various
correlation functions observed at
strong coupling should also be present
at small $\lam$. Such structure include
the signatures of black hole
singularities~\refs{\shenker,\guidoliu}
and the bulk-cone
singularities~\refs{\hubeny}.

Given that an arrow of time emerges for
small $\lam$ in the large $N$ limit, it
is natural to ask what should be the
string theory interpretation of the
high temperature phase for $\NN=4$ SYM
on $S^3$ at weak coupling, or from the
microcanonical point of view, what is
the bulk interpretation for a generic
state in the high energy sector.

{}From the parameter relations in AdS/CFT,
 \eqn\tieap{
 {l_s^2 \ov  R^2 } = {1 \ov  \sqrt{\lam}},
\qquad {G_N \ov R^8} = { 1 \ov N^{2}},  \qquad G_N = l_p^8 \sim
g_s^2 l_s^8, \qquad  \ .
 }
one might conclude that at weak
coupling $\lam \ll 1$, $l_s \gg R$ ,
i.e. the string length $l_s$ is much
bigger than the AdS curvature radius
$R$. However, it seems unlikely one can
give an invariant meaning to the
statement. For example, even starting
with a metric with $R \ll l_s$, one
could perform a field redefinition of
the form $g_{\mu \nu} \to g_{\mu \nu} +
\apr R_{\mu \nu} + \cdots$. In terms of
new metric one then has $R \sim l_s$.
Thus it seems to us that even for $\lam
\ll 1$, the corresponding bulk string
theory should describe a spacetime of
stringy scale, rather than sub-stringy
scale. This is also expected from the
gauge theory point of view. At weak
coupling the only mass scale is the
inverse radius of the sphere and there
are no other lighter degrees of
freedom. Thus the string scale has to
be of the same order as that of the AdS
curvature scale.

Can one interpret the bulk configuration corresponding to the high
temperature phase at weak coupling as a stringy black hole? It
seems to us the answer is likely to be yes. Let us list the
properties that the corresponding bulk configuration should
satisfy as expected from gauge theory, assuming there is no
further large $N$ ``phase transition'' between weak and strong
couplings:

\item{1.}  The bulk configuration should have an entropy and free
energy of order $O(1/g_s^2)$.

\item{2.} The object absorbs all fundamental probes (since
boundary correlation functions decay with time).

\item{3.} The bulk geometry should have a horizon (since the
boundary theory has a continuous spectrum).

\item{4.} The bulk configuration is likely to have singularities
(since the signatures of the black hole singularities in gauge
theory at strong coupling cannot disappear as the coupling is
changed if there is no phase transition).

\item{5.}A generic matter distribution will collapse into such a
configuration (since in the boundary
theory, a generic initial state will
approach the thermal equilibrium).

\item{6.} Results in~\refs{\minW,\sundborg} indicate that the
Euclidean time circle in the dual geometry for the theory in the
high temperature phase should become contractible\foot{Note that
this alone cannot imply that the bulk geometry is a black hole
since even at zero coupling the time circle becomes contractible.
As we argued earlier in this paper real-time correlation functions
in free theory do not behave like those of a black hole.}.

\ndt From the properties above, it
seems appropriate to call it a stringy
black hole.

Finally let us mention that it is possible that a stronger version
of equation \ehr\ holds, i.e. for two generic states $\ket{i}$,
$\ket{j}$ in the high energy sector,
 \eqn\fjjf{
 \rho_{ij} = {1 \ov \Om (E)} A(\om; E)  \RR_{ij},
 }
 with
$$
 E = {E_i + E_j \ov 2}, \qquad \om = E_i - E_j \
 $$
and $\RR_{ij}$ a random matrix. Equation \fjjf\ is considered to
be the hallmark of  quantum chaos~\refs{\peres,\srednicki}\foot{It
has also been argued in~\refs{\srednicki} that if \fjjf\ holds,
then thermalization always occurs.}. Thus it is possible that
$\NN=4$ SYM is chaotic in the high energy sector\foot{Chaos in a
classical Yang-Mills theory was discussed before
in~\refs{\BiroSH}. Possible pole of quantum chaos in AdS/CFT and
in black hole physics has also been discussed before
in~\refs{\KalyanaRamaZJ,\SrednickiFE}.}. Such a chaotic behavior,
if it exists, might be related to the BKL behavior near a
spacelike singularity~\refs{\bkl}.

\bigskip
\noindent{\bf Acknowledgments}

We would like to thank O.~Aharony, T.~Banks, W.~Fischler,
S.~Hartnoll, V.~Hubeny, A.~Lawrence, J.~McGreevy, M.~Rangamani,
A.~Scardicchio, S.~Shenker, M.~Srednicki, L.~Susskind for useful
conversations. This work is supported in part by Alfred~P.~Sloan
Foundation and U.S. Department of Energy (D.O.E) OJI grant, and
funds provided by the U.S. Department of Energy (D.O.E) under
cooperative research agreement \#DF-FC02-94ER40818.

\appendix{A}{Parameter relations in AdS/CFT}

${\cal N}=4$ super Yang-Mills theory is
a conformally invariant theory with two
parameters: the rank of gauge group $N$
and the 't Hooft coupling $\lambda =
g_{\rm YM}^2 N$. From the
operator-state correspondence, physical
states of the theory on $S^3$ can be
obtained by acting with gauge invariant
operators on the vacuum and their
energies are given by the conformal
dimensions of the corresponding
operators.

It was conjectured in~\refs{\MaldacenaRE} that ${\cal N}=4$ SYM
gives a nonperturbative description of type IIB superstring theory
in $AdS_5 \times S_5$. The AdS string theory also has two
parameters: the ratio between string length $l_s$ and the
curvature radius $R$ of AdS, and the ratio between the (10d)
planck length $l_p$ and $R$. These ratios respectively
characterize classical stringy corrections and quantum
gravitational corrections beyond the classical supergravity. For
small $l_s/R$ and $l_p/R$, parameters of SYM theory and bulk
string theory are related by\foot{We omit order one numerical
constants.}~\refs{\maldat}
 \eqn\tieap{
 {\apr \ov  R^2 } = {1 \ov \sqrt{\lam}},
\qquad {G_N \ov R^8} = { 1 \ov N^{2}},  \qquad G_N = l_p^8 \sim
g_s^2 \apr^4 \ .
 }
The above relations indicate that the
classical supergravity limit is given
by the large $N$ and large $\lam$ limit
of the SYM theory. In particular, a
departure from the large $N$ limit of
the Yang-Mills theory corresponds to
turning on {\it quantum} gravitational
corrections in the AdS spacetime, while
a departure from the large $\lam$ limit
(with $N=\infty$) corresponds to
turning on {\it classical} stringy
corrections.

AdS/CFT implies an isomorphism between
the Hilbert space of the two theories.
In particular, any bulk configuration
with asymptotic $AdS_5$ boundary
conditions can be associated with a
state (pure or density matrix) of the
Yang-Mills theory. The mass $M$ of the
bulk configuration is related to the
energy $E$ in YM theory
as~\refs{\gkp,\witten}
 \eqn\ejrn{
 E \sim MR \ .
 }
Depending on how $E$ scales with $N$ in the large $N$ limit,
states of Yang-Mills theory are related to different objects in
string theory in AdS. For example those whose $E$ do not scale
with $N$ (i.e. of order $O(1)$) should correspond to fundamental
string states. An object in AdS with a classical mass $M$
satisfies
 \eqn\Gaks{
 G_N M = {\rm fixed},  \qquad G_N/R^8 \to 0
 }
{}From \tieap\ and \ejrn\ the corresponding state in YM theory
should have $E \sim O (N^2)$.

\appendix{B}{Self-energy in the real time formalism}

In this appendix we first review some
basic properties of real-time
correlation functions. We then prove
that the spectral density functions of
fundamental fields in \onedAc\ have a
discrete spectrum after the resummation
of the self-energy diagrams \`a la
Dyson.

\subsec{Analytic properties of various real-time functions}

Various real-time thermal Wightman function for an operator $\OO$
are defined by
 \eqn\GontP{\eqalign{
 G_{+} (t) & = {1 \ov Z} \Tr \le(e^{-\beta H} \OO (t) \OO (0)
 \ri) - C \cr
 G_{-} (t) & = {1 \ov Z} \Tr \le(e^{-\beta H} \OO (0) \OO (t)
 \ri) - C \cr
 G_F (t) & = \th (t) G_+ (t) + \th(-t) G_- (t),  \cr
 G_{R} (t) & = i \th (t) {1 \ov Z} \Tr \le(e^{-\beta H} [\OO (t), \OO (0)]
 \ri), \cr
 G_{A} (t) & = - i \th (-t) {1 \ov Z} \Tr \le(e^{-\beta H} [\OO (t), \OO
 (0)]
 \ri) \
 }}
where $Z$ is the partition function and  $C$ is a constant to be
specified below. It is also convenient to introduce
 \eqn\Gontt{\eqalign{
 G_{12} (t)
 = G_+ (t-i \beta/2)  \  \cr
 }}
which can be obtained from $G_+ (t)$ by
an analytic continuation.

By inserting complete sets of states in \GontP, $G_+ (t)$ can be
written as
 \eqn\defGp{\eqalign{
 G_+ (t)  & ={1 \ov Z}  \sum_{i \neq j}   e^{-i E_j t} e^{i E_i (t + i\beta)} \,
 \rho_{ij} \cr
 }}
where $i,j$ sum over the physical states of the theory and
$\rho_{ij} = |\vev{i|\OO(0)|j}|^2$. Comparing \defGp\ and \GontP,
$C$ is chosen to be
 \eqn\choC{
 C=  {1 \ov Z} \sum_{i}  e^{- E_i \beta} \,
 \rho_{ii}
 }
Note that $C$ is chosen so that the Fourier transform of $G_+ (t)$
does not have a ``contact'' term proportional to $\delta (\om)$.
Assuming the convergence of the sums is controlled by the
exponentials, it follows from
\defGp\ that $G_+(t)$ is analytic in $t$ within the range $ -\beta < {\rm
Im} \, t < 0$. Similarly $G_- (t)$ is analytic for $0 < {\rm Im}
\, t < \beta$ and $G_{12} (t)$ for $-{\beta \ov 2 } < {\rm Im} t <
{\beta \ov 2 }$.

Introducing the spectral density function
 \eqn\speDe{\eqalign{
 \rho (\om) & = (1- e^{-\beta \om} ) \,
 \sum_{i,j} (2 \pi) \delta
 (\om- E_i + E_j)  e^{-\beta E_j} \rho_{ij}
 }}
then the Fourier transforms of \GontP\
can be written as
 \eqn\speR{\eqalign{
  G_+ (\om) & =
 {1 \ov 1- e^{-\beta \om}} \rho (\om) \cr
  G_{12} (\om) & = e^{-\ha \beta \om} G_+ (\om) =  e^{\ha \beta \om} G_- (\om)
  = {1 \ov 2 \sinh {\beta \om \ov 2}} \rho (\om) \cr
  G_R (\om) & = -  \int_{-\infty}^\infty {d \om' \ov 2 \pi}
 {\rho (\om') \ov  \om - \om' + i \ep} \cr
  G_A (\om) & = -  \int_{-\infty}^\infty {d \om' \ov 2 \pi}
 {\rho (\om') \ov  \om - \om' -i \ep} \cr
  G_F (\om)  & = G_R (\om) + i  G_- (\om) \cr
 }}
{}From \speR\ we also have
 \eqn\fjep{
 \rho (\om) = - i (G_R (\om) - G_A (\om))
 }
We also note that the Euclidean correlation function in momentum
space can be obtained from
  \eqn\EUC{
  G_E (\om_l) = \cases{G_R (i \om_l) & $l \geq 0$ \cr\cr
                        G_A (i \om_l) & $l < 0$ \cr},
  \qquad \om_l = {2 \pi l \ov \beta}, \quad l \in \IZ
  }

Some further remarks:

\item{1.} From \speDe--\speR,
 \eqn\jfjs{
 \rho (-\om) = - \rho (\om), \qquad G_{12} (-\om) = G_{12} (\om),
 \qquad G_R (-\om) = G_A (\om) \ .
 }

\item{2.} For a theory with a discrete spectrum, from \speDe, the
spectral function $\rho (\om)$ and $G_+ (\om)$ are given by a sum
of discrete delta functions supported on the real axis, while $G_R
(\om)$ is given by a discrete sum of poles along the real axis.

\subsec{Self-energy in real-time formalism}

In this section we consider real-time
correlation functions of fundamental
fields $M_\al$ in perturbation theory
using the real-time formalism. We
denote various quantities in \GontP\
with $\OO = M_\al$ by $D_+^{(\al)},
D_F^{(\al)}$ etc and will suppress
superscript $\al$ from now on. We prove
that the corresponding spectral density
functions have a discrete spectrum
after the resummation of the
self-energy diagrams \`a la Dyson. For
simplicity, we will consider the high
temperature limit so that we can ignore
the singlet projection~(see \fjd).

In the real time formalism \semenoff\
the degrees of freedom of the theory
get doubled (see also~\refs{\bellac}).
For each original field (type 1) in
\onedAc\ one introduces an equivalent
field (type 2)\foot{In a path integral
derivation these correspond to the
fields whose time argument is $t- i
\sig$ and we will take $\sig = {\beta
\ov 2}$.} whose interaction vertices
differ by a sign from the ones for
fields of type 1. Vertices therefore do
not mix the two different kind of
fields but propagators do and are
written as a $2 \times 2$ matrix.  For
example, in frequency space the
propagator for $M_\al$ (in the
interacting theory) can be written as
$D_{ab} (\om), a, b = 1,2$ with each
component given by
 \eqn\fhsd{\eqalign{
 D_{11} (\om) & = D_F (\om), \qquad D_{22} (\om)  = D_{11}^{*}
 (-\om)\cr
 D_{12} (\om) &  = {e^{{\beta \ov 2}
 \om}\ov e^{\beta \om} -1} \rho (\om) , \qquad
 D_{21} (\om)  = D_{12} (\om)
 }}
$D_{ab}$ can be diagonalized as
 \eqn\rjsm{\eqalign{
D_{ab} & = U \pmatrix{D_g (\om) & 0 \cr 0 & D_g^* (\om) \cr} U \cr
 }}
with
 \eqn\udna{
U = \pmatrix{\cosh \ga & \sinh \ga \cr \sinh \ga & \cosh \ga \cr},
\qquad \cosh \ga = {e^{{\beta \ov 2} |\om|} \ov \sqrt{e^{\beta
|\om|} -1}}, \qquad \sinh \ga = {1 \ov \sqrt{e^{\beta |\om|} -1}}
 }
and \eqn\eucl{
 D_g (\om)=i \int {d \om'\ov 2\pi} {\rho(\om')\ov
\om-\om' + i \ep \om } =
 \cases{-i D_R (\om) & $\om > 0$ \cr\cr
        - i D_A (\om) & $\om < 0$ \cr
 }}
The last expression in \eucl\ implies
that when analytically continued from
the positive real axis, $D_g (\om)$
cannot have singularities in the upper
half $\om$-plane. Similarly when
analytically continued from the
negative real axis, $D_g (\om)$ cannot
have singularities in the lower half
$\om$-plane. Note that $D_g (\om)$ can
have a discontinuity at $Im(\om)=0$. If
$D_g (\om)$ does turn out to be
analytic on the real axis, then it can
have singularities only on the real
axis in the limit $ \ep \to 0$, which
in turn implies that $D_R, D_A$ and
$D_F$ can have singularities only on
the real axis in the limit $ \ep \to
0$.

We will now show that $D_g (\om)$
obtained using the Dyson equation from
any finite order computation of the
self-energy is a rational function with
singularities only on the real axis.
This implies that the spectral function
$\rho$ consists of a sum of finite
number of delta functions supported on
the real axis.

Note that the Dyson equation can be written as
 \eqn\hana{
 {1 \ov D_g (\om)} =
  {1 \ov D_g^{(0)}(\om)} - i \tilde{\Pi}(\om)
 }
where
 \eqn\fjsp{
 D_g^{(0)} =
{i \ov \om^2 - m^2 + i \ep}
 }
is the free theory expression and $\tilde \Pi (\om)$ can be
computed from perturbation theory as follows: (i) Compute $2
\times 2$ matrix $\Pi_{ab}(\om)$ from the sum of amputated 1PI
diagrams for the propagator in real time formalism; (ii)
Diagonalizing $\Pi_{ab} (\om)$ using \udna, i.e.
 \eqn\Pjsm{\eqalign{
 \Pi_{ab} & = U \pmatrix{\tilde \Pi (\om) & 0 \cr 0 & \tilde \Pi^* (\om) \cr}
 U \ .
 }}
That $\Pi_{ab}$ can be diagonalized using $U$ is a consequence
\rjsm.

Now expanding $D_g$ and $\tilde \Pi$ in power series of $\lam$
 \eqn\abdp{\eqalign{
 D_g & = D_g^{(0)} + \lam D_g^{(1)} + \lam^2 D_g^{(2)} + \cdots \cr
 \tilde \Pi & =  \lam \tilde \Pi^{(1)} + \lam^2 \tilde \Pi^{(2)} + \cdots
 }}
from equation \hana\ we have
 \eqn\rnsK{
 D_g^{(1)} = D_g^{(0)} (i \tilde \Pi^{(1)}) D_g^{(0)} , \qquad
 D_g^{(2)} =D_g^{(0)} (i \tilde \Pi^{(2)}) D_g^{(0)} + D_g^{(0)} (i \tilde \Pi^{(1)}) D_g^{(0)}
 (i \tilde \Pi^{(1)}) D_g^{(0)}
 , \quad \cdots
 }
>From our discussion in section 3
(applied to fundamental fields), at any
finite order in perturbation theory
$\rho (\om)$ consists of sums of terms
of the form \rjeo. Plugging such a
$\rho (\om)$ into \eucl\ one finds that
$D_g^{(n)} (\om)$ is a rational
function and is analytic on the real
axis at each order in the perturbative
expansion (i.e. there is no
discontinuity at $Im(\om)=0$). Using
\rnsK\ we find that $\tilde \Pi^{(n)}
(\om)$ must also be a rational function
and analytic on real axis. This in turn
implies that the resummed $D_g (\om)$
found from \hana\ is a rational
function and analytic on real axis. We
conclude that the singularities of
$D_g$ must lie on the real axis and
there are only a finite number of them
at any finite order in the computation
of the self-energy $\Pi$. From \eucl\
the spectral density function must be a
finite sum of delta functions supported
on the real axis.

\appendix{C}{Energy spectrum and eigenvectors of sparse random
matrices}

In this appendix we summarize features of eigenvalues and
eigenvectors of a random sparse matrix found
in~\refs{\Rodgers,\Fyodorov,\Semerjian}. Consider an $M \times M$
real symmetric matrix $A$ whose elements $A_{ij}$ for $i\geq j$
are independent identically distributed random variables with even
probability distribution $f(A_{ij})$. Let $f(x)$ be of the
following form: \eqn\formdistr{f(x)=(1-\alpha)\delta(x)+\alpha
h(x)} where $0<\alpha<1$ and $h(x)$ is even and not delta-function
like at $x=0$. Let the variance of $h(x)$ be $v^2$. The parameter
$\alpha$ measures the sparsity of the matrix: for each row or
column of the matrix there will be on average $\alpha M=K$
elements which are different from zero. $K$ is called the
connectivity of the matrix. When $K< 1$ it is possible for
eigenvectors to be localized in a subspace with dimension smaller
than $M$. For $K>1$ in the large $M$ limit no such localization
occurs and the matrix has to be diagonalized in the full $M$
dimensional space.

When $K \gg 1$, the density of states reduces to Wigner's
semicircular law in an expansion in $K^{-1}$:
 \eqn\largeK{
 \rho(E)={1\ov 2 \pi \Gamma^2}\sqrt{4
\Gamma^2-E^2}\le(1+O(K^{-1}\ri))
 }
 Where $E$ is the eigenvalue value and $\Gamma$ is given by:
  \eqn\gam{\Gamma^2= K v^2\le(1+O(K^{-1})\ri)
 }
Notice that $Kv^2$ is the average value of
 \eqn\djer{
 \Gamma_i^2=\sum_{j\neq i} |A_{ij}|^2
 }
  over the rows or columns of the sparse matrix. The
first correction to $\rho(E)$ gives a change in the edge location,
however there also are nonperturbative tails to the distribution
which for $E\gg \Gamma$ assume the form:\eqn\npert{\rho(E)\sim
\le({E^2\ov e K}\ri)^{-E^2}} Their effect is to make the spectrum
unbounded.

Denote by $T$ the orthogonal change of basis matrix which brings
$A$ to diagonal form for $K\gg 1$. $T$ has a random uniform
distribution over the group of orthogonal $M\times M$ matrices.
Therefore the eigenvectors of $A$  are a random orthonormal basis
of the total space which means that apart from
correlations\foot{which are due to normalization conditions.}
which are negligible in the large $M$ limit their elements are
independently distributed gaussian random variables with mean $0$
and variance ${1\ov M}$. In particular the eigenvectors are
completely delocalized. Therefore for large $K$ the situation is
similar to that for the Gaussian Orthogonal Ensemble (GOE).

\appendix{D}{Single anharmonic oscillator\foot{This section is motivated from a discussion
with Steve Shenker.}}


It is clear that the argument presented
in section 4 applies to the real time
correlation functions of a single
anharmonic oscillator at finite
temperature (with changes of
combinatorial factors)
 \eqn\sans{
 S = \int dt \, \le(\ha \dot x^2 - \ha  x^2 - {1 \ov 4!}
 \lam x^4 \ri) \ .
 }
For example one can conclude that the
perturbation theory for
 \eqn\ajd{
 D_+ (t) = \vev{x(t) x(0)}_{\beta}
 }
should diverge at a time scale \ejs\ for $T \gg \om_0$ (we set
$\om_0 = 1$ in \sans). Here we give an alternative derivation of
this. Inserting complete sets of states in \ajd\ we find that
 \eqn\azaa{
 D_+ (t) =
Z^{-1}\sum_{n,m}|\vev{n|x|m}|^2 e^{-\beta E_n -it (E_m-E_n)}
 }
where $Z=\sum_n e^{-\beta E_n}$ and $\ket{n}$ are interacting
theory eigenstates. If we are interested only in contributions of
the form $(\lambda t)^n$ we get:
 \eqn\nns{
D_+ (t)=Z_0^{-1}\sum_{n,m}|\vev{n|x|m}_0|^2 e^{-\beta E^{(0)}_n -i
t(E^{(0)}_m-E^{(0)}_n)- it \lambda (E^{(1)}_m-E^{(1)}_n)} + \cdots
 }
where quantities with index $0$ are computed in the free theory
and $\lambda E^{(1)}$ are the energy shifts at first order in
perturbation theory. Equation \nns\ can be evaluated as
 \eqn\ressho{
 D_+ (t) =\ha Z_0^{-1}\sum_{n=0}^{\infty}(n+1) [e^{-\beta(n+\ha)
-it (1+{\lambda\ov 8}(n+1))}+e^{-\beta(n+{3\ov 2})
+it(1+{\lambda\ov 8}(n+1))}] + \cdots
 }
 which can be summed to give
 \eqn\resm{
 D_+(t)={(e^{\beta}-1)e^{\beta-it(1-{\lambda\ov8})}\ov
2(e^{\beta+{it \lambda\ov
8}}-1)^2}+{(e^{\beta}-1)e^{it(1-{\lambda\ov 8})}\ov 2(e^{\beta-{it
\lambda\ov 8}}-1)^2} + \cdots
 }
In \resm\ there are double poles at
 \eqn\rjgs{
 t=  \pm i {8 \beta\ov
\lambda}+ k {16 \pi \ov\lambda}, \qquad  k\in Z \ .
 }
If one resums the diagrams discussed in section 4, one would then
get simple poles and the positions of the poles are further away
from the real axis than those of \resm\ indicating that there are
some positive contributions not captured by the class of Feynman
diagrams.

The reason for the behavior \ressho--\rjgs\ can be attributed to
the fact that the first order energy shift behaves as
 \eqn\rjsj{
 \lam (E_{n+1}^{(1)} - E_n^{(1)}) \propto \lam n \ .
 }
Thus when $n$ is sufficiently large i.e. $n \sim {1 \ov \lam}$,
 perturbation theory breaks down due to level crossing. Also
note that the divergence of perturbation theory  at $t \sim {1 \ov
\lam T}$ has nothing to do with the standard argument of the
breakdown of perturbation theory by taking $\lam \to -\lam$.
Indeed the behavior here is due to a single class of diagrams not to the $n!$ growth of the number of diagrams.

We emphasize that while from the Feynman diagram point of view the
discussion for anharmonic oscillators is almost identical to that
for a matrix quantum mechanics (except that for matrix quantum
mechanics one restricts to planar diagrams), the underlying
physics for the breakdown of perturbation theory appears to be
very different:

\item{1.} In the example of a single anharmonic oscillator,
perturbation theory is asymptotic, i.e. the $n$-th order expansion
contains $n!$ independent diagrams. In contrast, in the planar
expansion of a matrix quantum mechanics, the number of Feynmann
diagrams grows only like a power in $n$. The class of planar
diagrams we identified gives rise to $O(n!)$ contribution (in
frequency space) at the $n$-th order. In the anharmonic oscillator
example, given that the perturbative expansion is already
divergent, one cannot really draw any clear conclusion from this
class of diagrams. For instance, its contribution could be
overwhelmed by those from $n!$ other diagrams. In contrast, in the
case of matrix quantum mechanics, the contribution from the
particular class of diagrams makes an otherwise convergent
perturbative expansion divergent. Given that the nature of
perturbation theory is very different between the single
anharmonic oscillator and the planar matrix quantum mechanics, one
should be very careful in drawing any conclusion when comparing
them. In particular, the fact that the anharmonic oscillator has a
discrete spectrum does {\it not} imply that in the matrix quantum
mechanics case, the divergence of the subclass of diagrams and
hence the breakdown of perturbation theory are {\it not} related
to a possible underlying quasi-continuous spectrum.

\item{2.} As indicated earlier in this appendix for a single
anharmonic oscillator, the divergent behavior of the class of
Feynman diagrams considered in section 4 should have to do with
with level mixing for states of energy $O(1/\lambda)$. Applying
the same technique to a matrix quantum mechanics, one again
expects  to relate the divergent behavior of the class of Feynman
diagrams to the mixing of energy levels which dominate the thermal
ensemble (i.e. with energy $O(N^2)$). More explicitly, let us
write \nns\ for the matrix case as
  \eqn\nnns{
D_+ (t)=Z_0^{-1}\sum_{n} e^{-\beta E^{(0)}_n} \sum_m
|\vev{n|M|m}_0|^2 e^{-i t(E^{(0)}_m-E^{(0)}_n)- it \lambda
(E^{(1)}_m-E^{(1)}_n)} + \cdots
 }
As we discussed in the main text, the sum over $m$ in the above
equation will involve an exponentially large number of states with
free theory energies ranging over of order $O(N)$. A naive
estimate of $E^{(1)}_m-E^{(1)}_n$ also gives order $O(N)$. Here
unfortunately the story appears to be rather complicated and it
appears it is not possible to extract a divergent time scale $1/\lam T$
from \nnns.

 In summary,
the Feynman diagram argument demonstrates the breakdown of
perturbation theory, but does not tell us why or how it breaks
down. It is certainly possible that completely different mixing
behaviors in the energy levels may be reflected similarly by
Feynman diagrams. In the anharmonic oscillator example discrete
levels mix, while in the matrix quantum mechanics a
quasi-continuous spectrum mixes. One must be very careful in
extrapolating the results for an anharmonic oscillator to a matrix
quantum mechanics.

\appendix{E}{Estimate of various quantities}

We now estimate \ekDk--\fgKk\ after averaging them over states of
similar energies. We will be interested in how these quantities
scale with $N$ in the large $N$ limit. An important property that
we will assume below for these averaged quantities is that they
are slow-varying functions of $\ep$ or $E$. In the large $N$
limit, we can then estimate them using the corresponding thermal
averages, which can in turn be expressed in terms of various
correlation functions at finite temperature. For example, the
thermal average of $\Sig_i$ is
 \eqn\riew{
 \hat \Sig (\beta) =  {1 \ov Z} \sum_i e^{-\beta E_i} \Sig_i
  = {1 \ov Z} \int dE \,  e^{-\beta E} \Om  (E) \, \Sig (E)
 }
where $\Sig (E)$ is the microcanonical average and $\Om (E) =
e^{S(E)}$ the density of states. Since $\Sig (E)$ is a
slow-varying function of $E$, we can perform a saddle point
approximation of the last expression, yielding
 \eqn\riil{
 \Sig (E) \approx \hat \Sig (\beta_E)  \le(1 + O(1/N^2) \ri)
 }
with $\beta_E$ determined by ${\p S (E) \ov \p E} = \beta_E$.
 Using the last equality of \fgKk\ we can write $\hat \Sig (\beta)$ as
 \eqn\fgKi{\eqalign{
  \hat \Sig (\beta) &
= {1 \ov Z} \sum_{i, j, i\neq j} e^{-\beta E_i} |\vev{i|V|j}|^2
\cr
 & = \vev{V(0) V(0)}_{\beta} 
 }}
where $\vev{V(0) V(0)}_{\beta}$ denotes the connected Wightman
function as defined by \defGp.  
{}From the standard large $N$ scaling argument 
\fgKi\ is of order $O(N^2)$ (recall that we include a factor of
$N$ in the definition of $V$). Thus unless \fgKi\ is zero at
leading order we conclude that $\Sig (E)$ can be written in a form
 \eqn\rimd{
 \Sig (E) = N^2 h(\lam, E/N^2)
 }
where $h (\lam, \mu)$ is a function
independent of $N$. An exactly parallel
argument can be applied to $\sig (\ep)$
in which case \fgKi\ is replaced by
expectation values in free theory and
thus we find that
 \eqn\dbrr{
 \sig (\ep) = N^2 \tilde h (\lam, \ep/N^2)
 }
for some function $\tilde h$.

As another example, let us look at the thermal average of \egKk,
 \eqn\owmk{
 {1 \ov Z} \sum_{i} e^{-\beta E_i} \bar \ep_i
 = {1 \ov Z} \int dE \, e^{-\beta E} \Om (E) \, \bar \ep (E) \approx
 \bar \ep (E_\beta)
 }
Using the last equality of \egKk, the left hand side of \owmk\ can
in turn be written as
 \eqn\rill{
  E_{\beta} - \vev{V}_{\beta}
 }
where $\vev{V}_{\beta}$ is the thermal
one-point function of $V$ in the
interacting theory and scales with $N$
as $O(N^2)$. Thus we can write
 \eqn\poew{
 \bar \ep (E) =  N^2 g(\lam, E/N^2)
 }
for some function $h$. An exactly
parallel argument yields
 \eqn\ruiu{
 \bar E (\ep) =  N^2 \tilde g (\lam, \ep/N^2) \ .
 }


To summarize, we find that the averaged values of $\Ga (\ep)$ and
$\Delta (E)$ are both of order $O(N)$ in the `t Hooft limit for
any nonzero $\lam$.
 Thus in the large $N$ limit, both the correlation length between
 interacting theory energy levels and the energy range that the free theory states are mixed
 under perturbation go to
infinity.

\appendix{F}{Some useful relations}

In this appendix we derive some important relations which will be
used in Appendix~G to derive the matrix elements of an operator
$\OO$ between generic states in the high energy sector.

\subsec{Density of states}

The conservation of states implies that the density of states $\Om
(E)$ of the full theory and $\Om_0 (\ep)$ of the free theory
should be related by
 \eqn\reOs{
 \Om (E) = \Om_0 (\bar \ep (E)) {d \bar \ep (E) \ov d E}
 }
which implies
$$
{1 \ov \Om (E)}{d \Om (E) \ov d E} =  {d \bar \ep (E) \ov d E} {1
\ov \Om_0 (\bar \ep (E))} {d \Om_0 \ov d \ep} \biggr|_{\bar
\ep(E)} + {{d^2 \bar \ep (E) \ov dE^2} \ov {d \bar \ep (E) \ov
dE}}
$$
In the large $N$ limit the second term in the above equation
should be of order $O(1/N^2)$.  Thus we find that
 \eqn\rhds{
\beta (E) = \beta_0 (\bar \ep(E)) {d \bar \ep (E) \ov d E}
 }
with
 \eqn\rkew{
 \beta (E) = {1 \ov \Om (E)}{d \Om (E) \ov d E}, \qquad
 \beta_0 (\ep) =  {1
\ov \Om_0 ( \ep)} {d \Om_0 \ov d \ep} \ .
 }
We also expect that
 \eqn\hrle{
\bar \ep (\bar E (\ep)) \approx \ep
 }
Note that all the above relations are valid only to leading order
in $N$.

\subsec{Properties of $\chi_E (\ep)$ and $\rho_\ep (E)$}

Consider the microcanonical average of \fjs\ and \fksm, which we
denote as  $\rho_\ep (E)$ and  $\chi_E (\ep)$ respectively.
 From \fjs\ and \fksm\ one should have
 \eqn\euhe{\eqalign{
 \rho_\ep (E) 
 = {\Om (E)  \ov \Om_0 (\ep)}
 \chi_E (\ep) \ .
 }}
{}From \ioep\ we should also have
 \eqn\noep{
  \int d \ep \, \chi_E ( \ep) =1
 }
and
 \eqn\Noep{
 \int dE \, \rho_{\ep} (E) = {1 \ov \Om_0 (\ep) } \int dE \, {\Om (E) }
 \, \chi_E (\ep) = 1 \ .
 }

Given that
 \eqn\fjfs{
 \bar \ep (E) = \int d\ep \, \ep \, \chi_E (\ep) , \qquad
  \Sig (E) = \Delta^2 (E) = \int d \ep \, (\ep - \bar \ep (E))^2 \chi_E (\ep)
  }
we can write $\chi_E (\ep)$ as
 \eqn\rje{
\chi_E (\ep) = f_E (\ep - \bar \ep (E))
 }
with $f_E$ a function which has a spread of $\Delta (E) \sim
O(N)$. Since we expect $f_E (\om)$ to fall off quickly to zero in
the large $N$ limit outside the range $(-\ha \De(E), \ha \De
(E))$, equations \noep\ and \Noep\ lead to
 \eqn\uekm{
 \int_{-\infty}^\infty d \om \, f_E (\om) =1
 }
 and
 \eqn\iuep{
 \int_{-\infty}^\infty dE \, {\Om (E) \ov \Om_0 (\ep) }
 f_E (\ep - \bar \ep (E)) = 1
 }
Changing the integration variable of \iuep\ to $\ep'=\bar \ep (E)$
and using \hrle, we find that
 \eqn\ehhn{
 \int d \ep' \, {\Om_0 (\ep') \ov
 \Om_0 (\ep) } \, f_{\bar E (\ep')} (\ep - \ep') =
 \int_{-\infty}^\infty d \om \, f_{\bar E (\ep)} (\om) {\Om_0 (\ep-\om) \ov \Om_0 (\ep)} =1
 }
where in the second expression we have
replaced $f_{\bar E (\ep')}$ by
$f_{\bar E (\ep)}$. This is because, as
a function of $\ep - \ep'$, the spread
of $f$ is of order $O(N)$, while $\bar
E (\ep') \approx \bar E (\ep) + O({\ep'
- \ep \ov N^2}) \approx \bar E (\ep)$.
The second expression of \ehhn\ can now
be written as
 \eqn\rhjj{
 \int_{-\infty}^\infty d \om f_E (\om) e^{-\beta_0 (\bar \ep (E)) \om} =1
 }
Equations \uekm\ and \rhjj\ can be written in a more symmetric
manner as
 \eqn\jndk{
 \int_{-\infty}^\infty d \om \, e^{\ha \beta (E) \om} \, g_E (\om)
 =  \int_{-\infty}^\infty d \om \, e^{-\ha \beta (E) \om} \, g_E (\om) = 1
 }
where we have introduced a function
 \eqn\newR{
 g_E (\om) = e^{-\ha \beta (E) \om} {d \bar \ep (E) \ov dE} f_E
 \le({d \bar \ep (E) \ov dE} \om \ri) \ .
 }
Equations \jndk\ imply that $g_E (\om)$ should fall off faster
than $e^{-\ha \beta (E) |\om|}$ as $\om \to \pm \infty$.

\subsec{A relation between matrix elements and correlation
functions in free theory}

In this subsection we derive in free theory a relation between the
matrix elements of an operator $\OO$ between states in the high
energy sector and correlation functions. For simplicity we
consider theories with a single fundamental frequency $\om_0$,
like $\NN=4$ SYM or \GonedAc.

The Lehmann spectral decomposition for frequency space Wightman
function $G_+^{(0)} (\om)$ of some operator $\OO$ in free theory
can be written as
 \eqn\foro{
G_+^{(0)} (\om) = {1 \ov Z_0} \sum_{a,b} e^{-\beta \ep_a}
\rho_{ab} \, \delta (\om -\ep_b +\ep_a)
 }
where $\rho_{ab} =
|\vev{a|\OO(0)|b}|^2$. Due to energy
conservation, $\OO$ can only connect
levels whose energy differences lie
between $-\Delta \om_0$ and $\Delta
\om_0$, where $\Delta$ is the dimension
of $\OO$, i.e. $\rho_{ab}$ can only be
non-vanishing for $|\ep_a - \ep_b| \leq
\Delta \om_0$. We can thus rewrite
\foro\ as
 \eqn\erid{
 G_+^{(0)} (\om) =  {1 \ov Z_0} \sum_{k=-\Delta}^\Delta G_k \delta (\om -k \om_0)
 }
with
 \eqn\rusp{\eqalign{ G_k & = {1 \ov Z_0} \sum_{a} e^{-\beta \ep_a}
\sum_{\ep_b = \ep_a + k \om_0} \rho_{ab}
 \cr
 & = {1 \ov Z_0}\sum_{a} e^{-\beta \ep_a}  \rho_k (a) \cr
 }}
 where $\sum_{\ep_b = \ep}$ denotes that one sums
over $\ket{b}$ whose energy is given by $\ep_b = \ep$. Note here
we have assumed that the free theory energy levels are equally
spaced as in $\NN=4$ SYM theory on $S^3$. We also introduced
 \eqn\risp{
 \rho_k (a) = \sum_{\ep_b = \ep_a + k \om_0} \rho_{ab}
 }
We now separate the sum $a$ in \rusp\ in terms of energies and
degeneracies, i.e.
$$
\sum_a = \sum_{\ep} \sum_{\ep_a = \ep}
$$
We thus find that
 \eqn\fkso{\eqalign{
 G_k
 & = {1 \ov Z_0} \sum_{\ep} \NN(\ep)
e^{-\beta \ep} \bar \rho_k (\ep) \cr
 }}
 where we have introduced the micro-canonical average of $\rho_k
 (b)$for energy $\ep$
 \eqn\rijs{
\bar \rho_k (\ep) = {1 \ov \NN(\ep)} \sum_{a \in \ep} \rho_k (a)
 = {1 \ov \NN(\ep)} \sum_{\ep_a = \ep} \sum_{\ep_b = \ep +k \om_0} \rho_{ab}
 \ .
 }
We expect that the microcanonical average $\bar \rho_k (\ep)$
should be a slow varying function of $\ep$, i.e. it can be written
in a form $N^\al f (\ep/N^2)$ for some constant $\al$. In the
large $N$ limit since $\NN (\ep) e^{-\beta \ep}$ is sharply peaked
at $\ep_\beta$ specified by, one can perform a saddle point
approximation in \fkso\ to get
 \eqn\fjsn{
G_k = \bar \rho_k (\ep_\beta) + \cdots
 }
>From \foro\ we thus find that
 \eqn\djsq{
 G_+ (\om) = \sum_k \bar \rho_k (\ep_\beta) \delta (\om-k \om_0)
 }
In the large $N$ limit since the connected part of $G_+ (\om)$
scales with $N$ as $O(N^0)$, thus we find from \djsq\ that
 \eqn\diue{
 \bar \rho_k (\ep_\beta) \sim O(N^0) \ .
 }

\appendix{G}{Derivation of matrix elements}

In this appendix we give a derivation of \ehr.  The main object of
interests to us is
$$\eqalign{
 \rho_{ij} & = \sum_{a,b} |c_{ia}|^2 |c_{jb}|^2 \rho_{ab} \cr
 & = \sum_{a} |c_{ia}|^2 \sum_k \sum_{\ep_b = \ep_a + k \om_0} |c_{jb}|^2 \rho_{ab}
 }$$
Due to the sparse and random nature of
$\rho_{ab}$, one cannot naively
approximate the sums over $a$ and $b$
in by integrals. Instead one must be
careful with the discreteness nature of
the sum. Note that
$$
\sum_{\ep_b = \ep_a + k \om_0} |c_{jb}|^2 \rho_{ab} \approx \bar
c_j (\ep_a + k)
 \sum_{\ep_b = \ep_a + k \om_0} \rho_{ab} = \bar c_j (\ep_a + k) \rho_k (a)
$$
which can be justified as follows.
Inside a given energy shell,
$\rho_{ab}$ can be treated as a random
sparse matrix. Thus one can treat the
summand as a random sampling of
$|c_{jb}|^2$. Since the number of
sampling points goes to infinity (as a
power in $N$) in the large $N$ limit,
we can approximate $|c_{jb}|^2$ by its
average value of the energy shell. We
now have
 \eqn\orie{\eqalign{
 \rho_{ij} & = \sum_k \sum_{a} |c_{ia}|^2 \bar c_j (\ep_a + k)
 \rho_k (a) \cr
 & = \sum_k \sum_{\ep} \bar c_j (\ep + k) \sum_{\ep_a = \ep}
 |c_{ia}|^2 \rho_k (a) \cr
 & = \sum_k \sum_{\ep} \bar c_j (\ep + k) \NN (\ep) \bar c_i
 (\ep) \bar \rho_k (\ep) \cr
 }}
 In the second line above we separated the sum over all states $a$
 into the sum over the energy and the sum over states in each
 energy shell. In the third line we replaced the sum in an energy shell
by its average values. The replacement
is all right since $|c_{ia}|^2$ and
$\rho_k (a)$ are completely independent
variables, so the average of their
product should factorize.

Now given that all quantities in the
last line of \orie\ are averaged
quantities, we approximate the sum over
$\ep$ by an integral. Averaging $i,j$
over states of the same energy and
using \fksm, we find
\eqn\fbrk{\eqalign{ \rho_{E_1 E_2} & =
\sum_k \int {d \ep \ov \Om_0 (\ep +k)}
\, \chi_{E_2} (\ep + k) \chi_{E_1}
(\ep) \, \bar \rho_k (\ep) \cr
 & = \sum_k \int {d \ep \ov \Om_0 (\ep +k)} \,
f_{E_2} (\ep + k - \bar \ep_2 ) f_{E_1} (\ep - \bar \ep_1) \, \bar
\rho_k (\ep) \cr
 & = \sum_k \int {d p\ov \Om_0 (\bep_{12} +p+k)} \,
f_{E_2} (p + k + \ha \De_{12} ) f_{E_1} (p - \ha \De_{12}) \, \bar
\rho_k (p + \bar \ep_{12}) \cr
 }}
with
$$
\bep_{1,2} = \bep (E_{1,2}), \qquad \bep_{12} = \ha (\bep (E_{1})
+ \bep (E_{2})) = \bar \ep (E) , \qquad \De_{12} = \bep (E_{1}) -
\bep (E_{2}) = {d \bar \ep (E) \ov dE}\biggr|_{E} \om
$$
where
$$
 E = {E_1 + E_2 \ov 2}, \qquad \om = E_1 - E_2
 $$
Equation \fbrk\ can be further simplified as
 \eqn\hbrk{\eqalign{
 \rho_{E_1 E_2} & = {1 \ov \Om (E)} \sum_k G_{12} (k)
 \int_{-\infty}^\infty d p \, g_{E} (p + k' + \ha \om)
  \, g_{E} (p - \ha \om) \cr
   & = {1 \ov \Om (E)} A (\om; E)
  }}
with
 \eqn\rhj{
 G_{12} (k) = e^{-\ha \beta_0 (\bar \ep_{12}) k} \bar \rho_k (\bar
 \ep_{12}), \qquad k' = {1 \ov {d \bar \ep (E) \ov dE}} k
 }
and $g_E  (\om)$ was defined in \newR.
Note that from equation \djsq, $G_{12}$
are essentially the Fourier components
of free theory correlation functions.
$A (\om;E)$ should be a smooth function
of $\om$ since the integral in \hbrk\
appears to be well defined for all
$\om$. It is easy to check that
 \eqn\rjds{
 A (-\om; E) = A (\om; E)
 }
 since $G_{12} (k) = G_{12} (-k)$. Further as $\om \to \infty$,
 we find that
 \eqn\ej{
 A (\om; E) \propto e^{-\ha \beta (E) |\om|} \ .
 }

Now let us examine possible singularities of $A(\om,E)$ on the
real axis. We start with the definition \rje\ of $f_E$. Since
$\chi_E (\ep)$ is the average of \fksm\ over states of similar
energies, $f_E(\om)$ must be a real positive function of $\om\in
R$. Then the function $g_E (\om)$ introduced in \newR\ should also
be real and positive as $\bar \ep (E)$ is a monotonous
 function of~$E$. The positivity and normalization conditions \jndk\
 imply that $g_E(\om)$ can at most have integrable singularities of the
 form\foot{Such integrable singularities can
 only arise if  $c_{ia}$ have accumulation points in the large $N$
 limit. While it appears unlikely that this can happen, we do not have a rigorous
 proof at the moment.
 }
 \eqn\sinGg{
 g_E(\om) \approx {K_i\ov |\om-\om_i|^{\alpha_i}}, \qquad \om \to
 \om_i, \quad \al_i < 1
  }
 Note that the closer $\alpha_i$ is to one the
smaller is $K_i$ from the normalization requirement. Now let us
look at the definition \hbrk\ of $A(\om;E)$,
 \eqn\hbrka{\eqalign{
 A (\om; E) & =  \sum_k G_{12} (k) s (\om + k)
  }}
where
 \eqn\intsh{
 s(\om)=\int_{-\infty}^\infty d x \, g_{E} (x + \ha \om)
  \, g_{E} (x - \ha \om) \ .
  }
Note that the finite sum over $k$ in
\hbrka\ cannot introduce singularities
in $\om$ therefore we focus on
$s(\om)$.  As $g_E (\om)$ falls off
faster than $e^{-\ha \beta (E) |\om|}$
as $\om \to \pm \infty$, the integral
in \intsh\ is convergent for
$x\rightarrow \pm \infty$. Thus we only
need to worry about possible
divergences arising from the middle of
the integration range. Integrating
\intsh\ we find that
 \eqn\intsomp{
 \int_{-\infty}^{\infty} d\om
 s(\om)=\int_{-\infty}^{\infty}dx g_E(x)\int_{-\infty}^{\infty} dy g_E(y)
 }
 which is finite by \jndk\ .
 Therefore the only singularities
allowed for $s(\om)$ are of integrable
kind ${K\ov |\om-\om_s|^{\alpha}}$ with
$\alpha<1$. We can find the locations
of $\om_s$ in terms of (integrable)
singularities of $g_E (\om)$ as
follows. Since $g_E(x-\ha \om)$ and
$g_E(x+\ha \om)$ are both integrable
the only possible divergences of
\intsh\ are at values of $\om$ for
which the integrable singularities of
two function sit on top of each other.
This happens for $\om=\om_i-\om_j$
where the $\om_i$ are the locations of
the singularities for $g_E(\om)$. For
$\om=\om_i-\om_j+\ep$ with $\ep$ small
the integral \intsh\ near $x \approx
\ha(\om_i+\om_j)$ can be written as
 $K_iK_j\int_{-\delta}^{\delta} dy {1\ov|y-\ha
\ep|^{\alpha_j}|y+\ha \ep|^{\alpha_i}}$
 where $\delta$ is some multiple of $\ep$.
By rescaling we see that it behaves as
$\ep^{1-\alpha_i-\alpha_j}$. Therefore the integral $s(\om)$ can
at most have a singularity of the form ${K_iK_j\ov
|\om-\om_i+\om_j|^{\alpha}}$ with $\alpha=\alpha_i+\alpha_j-1<1$.

Thus we conclude that on the real axis $A(\om ;E)$  can have at
most integrable singularities of the form
 \eqn\ejjsP{
 A(\om; E) \propto {1 \ov |\om- \om_s|^{\al_s}}, \qquad \al_s < 1
 \ .
 }


\listrefs

\end
\end

For simplicity, we will assume that for a given $\ket{i}$,
$c_{ia}$ is supported inside a region of width $\Delta_i$ around
$\ep_i$ only. The possibility of small tails outside the region
will be considered later. Thus one has
$$
|c_{ia}|^2 = \cases{{1 \ov \Om (\ep_i) \Delta_i}, & $|\ep_a -
\ep_i| < \Delta_i$ \cr\cr
 0 & ${\rm otherwise}$ \cr
 }
$$
where $\Om (\ep)$ is the density of states of the unperturbed
theory. We will also

Can we show that when $\beta \to 1$ that $|\OO_{ij}|^2$ above has
the following form ?
$$
|\OO_{ij}|^2 = {1 \ov \Om (E_t)} A (\om;\beta), \qquad
$$
with
$$
 E_t = \ha (E_i + E_j), \qquad \om = E_i - E_j, \qquad \beta = {\p
 \log \Om(E_t) \ov \p E_t}
 $$

The theory is rather complicated theory with an infinite number of
matrices $M_a$ and $\xi_a$ of increasingly larger frequencies and
infinite number of interaction terms. {\bf Check UV divergence in
Yang-Mills theory.}

In the first part of this section we argue that this cannot be
captured by the standard perturbation theory at any finite order.
In the second part we discuss an alternative way to attack the
problem in a simplified version of Yang-Mills theory on $S^3$.

In this limit, given a generic state of energy
$E = cN^2$ in the Yang-Mills theory, the state should evolve
into a thermal density matrix and almost all the information (except those associated
with conserved quantum numbers) of the initial
state is lost. In the bulk language, a black
hole is formed.\foot{Note that the information loss is only a consequence of
taking the large $N$ limit. Black holes are not really black at finite
$N$, i.e. in a full non-perturbative quantum gravitational treatment.}

\item{1.} A free Yang-Mills theory on $S^3$
never thermalizes in the $N \to \infty$ limit.

\item{2.} We show that in the large $N$ limit, with any nonzero `t
Hooft coupling $\lam$, there are an exponentially (in $N^2$) large
number of free theory states within energy  ranges proportional
$N$ mix under the interaction. We argue as a consequence, the system almost
always thermalizes. ({\bf we also derive
some results on correlation functions.})

\item{3.} We show that the effect in (2) cannot be captured by the
planar perturbative expansion. We show for real-time correlation
functions at finite temperature, the planar perturbative expansion
has zero radius of convergence in the complex $\lam$-plane\foot{Recall
that since the number of planar diagrams grow with the order of perturbation
as a power, planar expansion for zero temperature correlation functions
has a finite radius of convergence in $\lam$.}. In
particular, the corresponding Lorentzian correlation functions in
frequency space have essential singularities at $\lam=0$.

The plan for the paper is as follows. In section 2 we formulate
more precisely the properties we would like to establish. In sec 3
we show that free YM theory on $S^3$ never thermalizes. In sec. 4
we argue that the system does thermalize for any nonzero $\lam$.
In sec. 5 we show that the planar expansion for real-time
correlation functions has zero radius of convergence. In sec. 6 we
discuss the interpretation of our results in the physics of black
hole.

If the theory under considerations have several incommensurate
fundamental frequencies, one simply includes a sum like those
\rjek\ and
 for each such frequency. The maximal number of independent
exponentials are $2^q$, where $q$ is the total number (independent
of $N$) of matrices in $\OO$.

{\bf Note that in AdS/CFT $\OO$ is dual to fundamental string
states, while in $\OO$ contain few number of degrees of freedom.}

Equation  is intrinsically non-perturbative. One can show that at
any order in perturbation theory spectral functions $\rho_{E_1
E_2}$ are supported at a finite number of discrete frequencies and
$G_R (t)$ are  oscillatory. We also show that the perturbative
expansion for $G_R (t)$ in the planar limit has a zero radius of
convergence in the $\lam$-plane\foot{Recall that in the planar
limit, zero temperature perturbation theory has a finite radius of
convergence in $\lam$.}. Thus perturbation theory does not capture
the long-time behavior of $G_R (t)$, for any nonzero $\lam$, no
matter how small. The reason for the breakdown of perturbation
theory and for  can be attributed to the following simple facts.
In the free theory limit $\lam =0$, states of energies $E = cN^2$
are highly degenerate, with degeneracies of order $O(e^{d N^2})$
for some number $d (c)$. When one turns on a nonzero $\lam$, no
matter how small, an exponentially large (in $N^2$) number of free
theory states mix under the interaction. In particular, simple
power counting shows that the range of energy of states which are
mixed is of order $O(N)$.

Another motivation for our study is to explore the connection
between the appearance of the black hole singularity in the bulk
and thermalization in the boundary YM theory\foot{See
also~\refs{\banks} for a very interesting recent discussion of
this.}. Our results give indication that the black hole
singularity in the string theory dual may survive the $\alpha'$
correction. In~\refs{\shenker,\guidoliu}, manifestations of the
black hole singularity in Yang-Mills theories were identified in
the large $N$ and large $\lam$ limit. Equations  are consistent
with general behavior of the strong coupling results found from
gravity. Unfortunately, the techniques of this paper do not give
us enough information on the analytic structure of $A_E (\om)$ to
be comparable to the manifestations found
in~\refs{\shenker,\guidoliu}. We leave such comparison for future
work.

\appendix{J}{Junk?}

This may be heuristically argued as follows. Since the number of
degrees of freedom of SYM theory is proportional to $N^2$, the
large $N$ limit of an SYM theory in a state whose energy scales
with $N$ as $E = cN^2$ is like a thermodynamic limit, {\it i.e.}
the number of degrees of freedom goes to infinity while the energy
per degree of freedom remains fixed. A basic hypothesis\foot{ Such
thermalization hypothesis, while widely believed to true, is
notoriously hard to prove.} of quantum statistical mechanics is
that an interacting many-body system almost always thermalizes
when sufficiently excited in the thermodynamic limit.

The simplest observables for this purpose is the real-time
correlation function in a canonical ensemble
 \eqn\ighn{
 G_+ (t) = {1 \ov Z} \th(t) \Tr \le(e^{-\beta H} \OO (t) \OO(0) \ri) , \qquad
 Z = \Tr e^{-\beta H} \
 }
We require operator to have dimension of order $O(1)$ and $\beta$
to be sufficiently small.

At finite $N$,

Direction of time:
 \eqn\fjjd{
 \lim_{t \to \infty} G_R (t) \to 0 \ .
 }
\fjjd\ indicates that under small perturbations, the system
eventually go back to thermal equilibrium in a canonical ensemble.

At strong coupling, behavior of $G_+$ can be obtained from gravity
and we find


